\documentclass[12pt]{article}		
\usepackage{fancyhdr}
\usepackage{multicol}
\usepackage{latexsym}
\usepackage{graphicx}
\usepackage[T1]{fontenc}			
\usepackage{setspace}				
\usepackage{endnotes}				
\usepackage{marvosym}				
\usepackage{seqsplit}				
\usepackage[hyphens]{url}

\usepackage[round,semicolon,authoryear]{natbib}	
\begin{document}
	
\title{AI, Trust, and Teaming: The Humans-as-Handlers Approach for Autonomous and Opaque AI Systems}
\author{Nathan Gabriel Wood		
	\\nathan.wood\MVAt tuhh.de}
\date{}

\maketitle	


\renewcommand*\abstractname{Abstract\hfill}
\begin{abstract}

Artificial intelligence (AI) is becoming ubiquitous, and across domains, increasingly autonomous systems are carrying out tasks which raise significant ethical and legal challenges which demonstrate a need for strong human-machine teams rooted in trust. In this article, I argue that within highly impactful areas (such as medicine or warfighting) there are grounds for us initially treating autonomous and opaque systems as relevantly analogous to dogs (or other animals with which we have close relationships). Under this analogy, humans making use of these systems are not to be viewed as ``users'' or ``deployers'' of these systems, but instead take the role of ``handlers''. This recasting of roles shifts the way we view humans, AI-enabled and autonomous systems, and the relations between them, and moreover clarifies the clear and traceable lines of responsibility humans have for the outcomes brought about when using these systems. In developing this point, I clarify that the machine-animal analogy does admit disanalogous elements, but that its touch-points ground it as a starting point. I then explore how we can divest the humans-as-handlers approach of those aspects of our relationships with animals which are unfitting for how we engage with and make use of autonomous and AI-enabled systems. I conclude by arguing that the trajectory of human-machine teamings for autonomous and AI-enabled systems should be a state where we authentically view these not as artifacts which we simply make use of, but as collaborators with which we pursue complex goals and carry out complex tasks. 

\bigskip
	
\noindent\textbf{Keywords:} \emph{Autonomous Systems, Artificial Intelligence, Trust, Opacity, Human-Machine Interaction, Autonomous Weapon Systems}		

\end{abstract}

\pagebreak

\section{Introduction}\label{sec_intro}

Artificial intelligence (AI) is becoming ubiquitous in modern society, from large language models (LLMs) streamlining everyday elements of business \citep{boiko2023emergent,fan2023bibliometric,hadi2023large,inagaki2023llms,williams2023algorithmic} to highly specialized AI systems being used in medical diagnostics \citep{johnson2021precision,rajpurkar2022ai,king2023future}. As AI enters into increasingly more domains, numerous ethical and legal challenges arise, ranging from copyright, data, and privacy concerns (notably, relating to the training of LLMs and image generation models \citep{stahl2018ethics,elliott2022ai,xiang2024fairness}) to deep-seated worries about ethically weighty decisions being outsourced to machines (e.g., most notably within the medical and military domains; see respectively \citep{anom2020ethics,morley2020ethics,gundersen2022future} and \citep{hrw2012losing,sparrow2016robots,roff2018trust,hrw2018heed}). Discussions of the impact of AI are further complicated by the fact that hype has caused many algorithmic processes to be touted as ``AI'', even when these do not have any seeming ``intelligence'' to speak of \citep{soni2020artificial,mattmann2024ai}. Yet even granting that some of the AI boom is (over-)hyped, it is clear that artificial intelligence is reshaping the way we carry out many tasks which are ethically and legally significant, and this alone provides some cause for worry. 

In order to address these concerns, regulatory bodies, academic institutions, and professional associations have put forward a variety of AI ethics principles, best practices, or technical solutions.\footnote{Notable governmental efforts are represented by the European Union's AI Act, and the words of the United States' Department of Defense regarding AI in the military domain. See, e.g., \citep{euhlegai2019ethics,euparliament2023ai,usdod2023directive3000.09}.} However, there still remain many pressing questions concerning how we may adequately team humans with autonomous and increasingly opaque AI-enabled systems. In societally impactful areas where lives will often be on the line (e.g., the medical and military domains), there is an especial need for robust methods of creating and sustaining strong human-machine teams, and for entrenching the trust needed for such teams to effectively, responsibly, and reliably function. 

In this article, I argue that within highly impactful areas such as medicine or warfighting, where mistakes are liable to quickly result in lost lives, one way for us to establish and sustain strong human-machine teams with high levels of trust is to look to the analogy of human-animal teams. In particular, I argue that in order to establish the needed trust for human-machine teams to effectively and responsibly function, we must recast how we conceptualize them; rather than viewing humans as ``users'' or ``deployers'' of such systems, we should instead view our role as that of ``handlers''. This shifting of language alters the assumptions surrounding what is expected of both humans and machines when they are jointly carrying out some task. Moreover, the humans-as-handlers approach clarifies the clear and traceable lines of responsibility humans have for the outcomes brought about through the use of such systems, helping to mitigate a particular set of concerns surrounding AI-enabled and autonomous systems (e.g., so-called ``responsibility gaps'').\footnote{See \citep{wood2023awsresponsibility,wood2024explainable}, responding to the problem as presented by, e.g., \citep{matthias2004responsibility,asaro2006should,sharkey2007automated,pagallo2011robots,wagner2014dehumanization,chengeta2016accountability,crootof2016war}.} In developing the arguments, I make clear that the machine-animal analogy does admit many disanalogous elements, but that its touch-points are significant enough to ground it as a starting point. I then explore how we can divest the humans-as-handlers approach of those aspects of our relationships with animals which are arguably unfitting for how we will or ought to engage with and make use of autonomous and AI-enabled systems. I conclude by arguing that the final trajectory of human-machine teamings for autonomous and AI-enabled systems in critical domains should be a state where we authentically view these systems not as artifacts which we simply make use of, but as collaborators with which we pursue complex goals and carry out complex tasks. 

I begin in Section \ref{sec_trust} by examining the concept of trust, making clear that while this notion is often invoked in a somewhat simplistic manner within AI ethics and the literature on AI regulation,\footnote{See, e.g., \citep{roff2018trust,euhlegai2019ethics,euparliament2023ai,ryan2020trust,omrani2022trust}. Cf., however, \citep{lyons2019trust,choung2022trust} for discussion closer to that defended here.} it is a far more nuanced concept. This is important to clarify at the outset, as any discussion of trust, transparency, and teaming between humans and machines must be precise about what sort of trust it has in mind. Moreover, the more abstract discussion of trust naturally leads to practical concerns regarding opacity in AI systems (Section \ref{sec_transparency}). In connecting trust and transparency to different aspects of human-machine teaming, we are further able to see how different facets of trust may be nurtured or diminished through different technical and institutional choices. Importantly, it is also made clear that changes to a system might simultaneously reduce one facet of the larger concept of ``trust'' while also enhancing some other facet. By canvassing the underlying concepts bound up in the notion of ``trust'', we are able to locate certain senses which are clearly linked to transparency and more fruitfully explore the underlying contours of trust and transparency as connected but highly distinct concepts. Section \ref{sec_handlerapproach} then presents the analogy of human-animal teams as a guide for modeling human-machine teams, highlighting disanalogies between the two models but defending a pragmatic and solution-oriented argument for taking the human-animal analogy as a starting point from which one may trim and expand details to fit the particularities of human-machine teams. Section \ref{sec_humansashandlers} then explores how the ``humans-as-handlers'' approach would function in practice, showing the benefits and risks of the approach. Furthermore, this section expounds and defends a \emph{collaborative} understanding of human-machine teams, where humans are not viewed as ``users'' or ``deployers'' of autonomous and AI-enabled systems, but rather where humans work together with these systems to reach a given set of goals. In Section \ref{sec_conclusion}, I conclude. 

Before moving onto the arguments themselves though, there are two points worth briefly addressing. First, it is worth noting that for every advanced technological system, there will almost certainly be a need to establish some degree of human-machine teaming, as virtually no system can be used without some level of training, familiarity, or simply even an understanding of its strengths and limitations. This is the case even for simpler systems, and indeed for mere artifacts such as handheld tools, as there are clearly some users who can better utilize these, and some whose use of them may be downright dangerous. For systems (and artifacts) used in societally low impact areas, rich human-machine teams may rarely be needed, as simple guidelines can give users enough of a handle on how to use these that they will be sufficiently competent. Moreover, the dangers posed by mistakes or faulty uses of a system are minimal for low impact systems.\footnote{For discussion of user guidelines to aid in the use of lower impact systems such as LLMs, see, e.g., \citep{barman2024beyond} and citations therein.} However, in highly impactful areas like medicine or defense, mistakes can be rather weighty, indeed lethal, and in those environments, trust and teaming cannot be boiled down to simple lists of ``dos and don'ts''. Rather, due to the significance of decisions made in these areas, and the regular need for speed in decision-making, trust and teaming take on a grave significance, and can be the difference between life and death for a patient or a warfighter. In what follows, our concern will be with autonomous and opaque AI-enabled systems in these domains, and the arguments should be understood as applying to, first and foremost, professionals in those domains. Specifically, we will focus on AI in defense, as military applications of AI represent the most difficult cases, given that time is generally very tightly constrained, life and death decisions are generally the norm, and soldiers will often have reduced (perhaps even little) ability to seek outside information or assistance in deciding how best to employ an AI-enabled or autonomous system. 

Second, given the centrality of ``artificial intelligence'' in what follows, it will be useful to give at least some definition of the term at the outset. Early AI research often set a high bar for AI, using definitions which pointed toward ``the same scope of intelligence as we see in human action: that in any real situation behavior appropriate to the ends of the system and adaptive to the demands of the environment can occur, within some limits of speed and complexity'' \citep[p. 116]{newell1976computer}, or which were ``concerned with methods of achieving goals in situations in which the information available has a certain complex character'' \citep[p. 308]{mccarthy1988mathematical}.\footnote{See also \citep{minsky1985society}.} Making this more precise, in what follows we will follow the definition of \citep{wang2019defining}, that artificial intelligence

\begin{quote}
	is the capacity of an information-processing system to adapt to its environment while operating with insufficient knowledge and resources \citep[p. 19]{wang2019defining}.\footnote{See also \citep{wang1995non} and the 2020 special issue of the \emph{Journal of Artificial General Intelligence} dedicated to discussing Wang's view (Volume 11, Issue 2). See also the EU AI Act, esp. p. 39.}
\end{quote}

Given that, under this definition, AI systems will inherently have some capacity to adapt to their environment, there is liable to always remain some degree of unpredictability and opacity in these. It is for this reason that discussions of transparency and trust are so important, as an adapting system making decisions based on incomplete information will rarely be wholly predictable. As such, it is incumbent upon the humans making use of such systems that they have the requisite understanding, sensitivity, and judgment to do so responsibly and effectively. This is all the more so relevant for AI used in such things as warfare, where mistakes or unpredictable actions can quickly lead to deaths, deaths which could be constitutive of war crimes if combatants are not acting with the due care required in war. 

\section{Shades of Trust}\label{sec_trust}

The literature on human-machine teaming and trust has a long pedigree,\footnote{See, e.g., \citep{oneill2020human} and citations therein.} and it is no surprise that trust has become one of the focal points in the debates currently being had at regulatory or legislative bodies. However, across scholarship and within governmental policy, trust is often treated as a flat concept with limited contours and a straightforward understanding. For example, the recent \emph{Artificial Intelligence Act} of European Parliament (hereafter, AI Act) discusses ``trustworthy artificial intelligence'' \citep[p. 160]{euparliament2023ai}, but provides no systematic understanding of trust, nor any analysis of what it means for individuals or societies to ``trust'' an AI system. Instead, reference is made to the 2019 \emph{Ethics Guidelines for Trustworthy AI} compiled by the European Commission's High-Level Expert Group on Artificial Intelligence, which talks of it being ``essential that trust remains the bedrock of societies'' \citep[p. 4]{euhlegai2019ethics} or that AI systems must be ``demonstrably worthy of trust'' (\emph{ibid}). A clear conception of trust is, however, only provided by brief reference to a single earlier work, without any analysis or further development:

\begin{quote}
	We take the following definition from the literature: ``Trust is viewed as: (1) a set of specific beliefs dealing with benevolence, competence, integrity, and predictability (trusting beliefs); (2) the willingness of one party to depend on another in a risky situation (trusting intention); or (3) the combination of these elements''(\citep[p. 38]{euhlegai2019ethics}, referencing \citep[p. 47]{siau2018building}).
\end{quote}

To properly grapple with the interactions between trust and transparency in AI systems, we must do more than simply stipulate some set of ``trusting beliefs'' and a vague notion of dependence in ``risky situations''. Rather, an understanding is necessary of how different aspects of trust are brought to the fore by different people, entities, or institutions, how those aspects may interact with one another, and what exact type of dependence (with or without risk) is entailed when we trust someone or something.\footnote{To be sure, \citep{euhlegai2019ethics} provides a rich and useful exploration of how we may strive toward trustworthy AI, with many helpful suggestions concerning design and implementation. However, throughout the discussion, there is virtually no exploration of how trust is understood by the humans who must ultimately place trust in AI systems, and this lack is problematic insofar as trust is a multi-faceted and deeply human concern which requires at least some exploration of how humans employ the term and what we mean by it in various contexts. Furthermore, in order to have a sufficiently incisive debate surrounding trust and transparency in AI systems, we must be sensitive to these underlying sentiments in order to craft best practices which relate to the various instantiations of ``trusting'' which humans may apply.} To help us explore these various points and in order to lay the necessary foundations for the humans-as-handlers approach developed in Sections \ref{sec_handlerapproach} and \ref{sec_humansashandlers}, we need a more nuanced view of trust, one which we may find by following the work of Bertram Malle and Daniel Ullman.

Philosophical analyses of trust often break down the concept into a tripartite relation between a trustor, a trustee, and action or state of affairs the trustor expects the trustee to carry out/bring about \citep{mayer1995integrative,hardin2002trust,baier2014trust}. The expected action or state of affairs may also imply some risk to the trustee, as included in the definition used by the European Commission.\footnote{For definitions including risk, see, e.g., \citep{koller1988risk,josang2004analysing}.} The core of these analyses is the idea of trust \emph{simpliciter}. However, even a casual survey of terms related to trust would indicate that there are many concepts which integrally connect to it and which may have bearing on how we view trust with regards to human-human interactions, human-animal interactions, and human-machine interactions. Moreover, closer examination of how we use a variety of terms relating to trust indicates that rather than a monolithic conception which aims to explicate only a single term and its usage, there are good grounds for us to, following \citep{malle2021multidimensional}, have a multidimensional measure of trust (MDMT).

Building on earlier works,\footnote{E.g., \citep{ullman2018what,ullman2019measuring}. For further development of Malle and Ullman's work, see also the more recent brief survey \citep{malle2023measuring}.} \cite{malle2021multidimensional} show that, beyond a simple measuring of trust, the many different words one may use to express a (dis)trusting attitude toward some person or object demonstrate that trust-related terms are grouped under a small number of concepts highlighting core aspects of ``trust''. In particular, they show that when faced with a question like, ``Can I trust this person?'', this often may not be a simple question inquiring into a single characteristic of someone, but rather a compressed shorthand for the more precise but cumbersome question, ``Is this person reliable, capable, sincere, and/or ethical?''. Through empirical analyses of how everyday people use and understand various trust-related terms, Malle and Ullman further show that these four subscales of trust (reliable, capable, sincere, ethical) are further superordinated into two larger categories, namely performance trust (reliable, capable) and moral trust (sincere, ethical). With machines, humans are more apt to be concerned with performance trust. On the other hand, human-human interactions usually include aspects of both performance trust and moral trust. However, despite the general tendency to view machines in a more performance-oriented manner, where we ``trust'' them to the extent they are capable, reliable, competent, consistent, etc.,\footnote{For the full list of terms captured under the trust groups of ``capable'' and ``reliable'', see figures 1 and 2 of \citep{malle2021multidimensional}, at p. 18.} there are cases when we may consider more ethical or ``human'' aspects of trust in our relations with machines. More than this, there are cases when we have good reason to examine these less performance-focused aspects of trust, and instead place weight on the moral characteristics of a machine. 

Consider an opaque AI used in war, say for weapons targeting purposes. We might ask ourselves, ``Is this AI trustworthy?'' or ``Can I trust this system?''. However, these simple questions mask the distinct facets of trust illuminated by Malle and Ullman's analysis, facets which may pull in different directions for a given system. Consider, for example, capability and reliability. Asking whether the AI is capable focuses on whether it can do some task. Asking whether it is reliable (plausibly) assumes that it has some capability to begin with, and so the question is about whether that capability is to be expected at all times and under a variety of conditions. Yet whether an AI system is capable and reliable in acquiring targets or guiding weapons may tell us nothing about the system's moral characteristics. This is important, as a system might capably and reliably locate enemy vehicles and guide weapons into these, but never take into account whether there are civilians nearby, and if so, how many. The presence of civilians and the likelihood of their deaths (so-called ``collateral damage'') is a critical moral and legal aspect of warfighting, and one which goes far beyond mere capacity or reliability in finding and striking targets. Quite on the contrary, the principle of proportionality will in many cases demand that one \emph{not strike}, even when weapons are fully capable of eliminating targets. And likewise, many of the moral and legal aspects of war call for abstaining from the use of weapons in particular circumstances. Thus, an AI system that is less reliable in \textit{striking} could be more ethical, if its unreliability in carrying out strikes is rooted in it having design features aimed at minimizing collateral damage. 

In any event, one system may more reliably and competently achieve some discrete goal, but achieve that competence and reliability at the cost of not showing restraint. Another system may be less competent and reliable from a purely strike-oriented perspective, but only because it has safeguards built in to make the system more moral or to show that the deployers of the system are sincere in their respect for the ethics and laws of war. If one wishes to ask, ``Which system is more trustworthy?'', the proper response should not be to give some rating to both systems along a single scale of ``trustworthiness''. Instead, we should rightly point out that such a question is inherently flawed, in that it demands we flatten a rich concept with many underlying points. Moreover, the question seems to implicitly assume that all systems are graded on the same scale, which simply is not the case. Rather, which system we trust often has as much to do with what we expect of them as it does with their technical achievements; a humanitarian peacekeeping force may find purely offensive and highly lethal drones to be deeply untrustworthy, simply in virtue of the peacekeepers' mission and the unlikelihood of such a drone aiding it, while a regular military unit engaged in peer-to-peer warfare might find the same drone highly trustworthy due to its capability and reliability in killing enemies. What makes one and the same drone (un)trustworthy may be the mission, not the drone. 

More broadly, it is critical that we not seek a simplified or flat understanding of trust when discussing human-machine teaming. Rather, we must be cognizant of the different ways we can trust, the different aspects of trust we can hone in on, and the ways separate facets of the concept may interact in one and the same system. Such a nuanced understanding will allow us to better discern what precisely undermines trust (in one of its facets, aspects, or characteristics) in particular cases, and help us to better entrench trust as a core value in teaming humans and opaque AI systems in morally and legally impactful domains. 

\section{Transparency and Trust}\label{sec_transparency}

So far we have focused primarily on trust and a broad idea of ``AI''. However, precision and clarity about what aspects of AI systems are thought to be morally and legally troubling further shows the importance of trust for overcoming these challenges. More than this, some of the problems raised by emerging AI technologies, many of which likely cannot be wholly addressed through technical solutions, highlight the need for a trust-based framework of training and use for societally impactful AI-enabled systems. 

If we recall the definition of artificial intelligence presented in Section \ref{sec_intro}, namely the ``capacity of an information-processing system to adapt to its environment while operating with insufficient knowledge and resources'' \citep[p. 19]{wang2019defining}, it should be clear that in the military domain, there will be enormous pressures to develop systems with ever greater degrees of artificial intelligence.\footnote{There are a variety of reasons for use of AI to expand across many other domains as well, but within the military the advantages offered by AI fit particularly cleanly with the demands of the discipline, especially as information-gathering apparatuses continue to improve and we continue to have ever greater amounts of data with which to make decisions in war, necessitating automation of certain tasks in order for humans to be able to engage at all with the new age of data-driven warfare.} This is because warfare is always conducted against a backdrop of limited knowledge and resources, and one's ability to find advantage given that reality is a central aspect of victory in arms. Yet the pressure to create increasingly advanced AI systems is likely to lead us toward systems which are increasingly opaque as well, and opacity raises a number of concerns. 

Simply put, ``as AI systems become more complex, it becomes increasingly difficult for humans to be able to fully comprehend, understand, or explain how they function'' \citep[p. 4]{wood2024explainable}, to the point where we risk arriving at a state where AI systems are ``a `black box'... for which we know the inputs and outputs but can't see the process by which [the system] turns the former into the latter'' \citep[p. iii]{michel2020black}. Across domains of application, opaque AI systems raise challenges regarding their responsible and effective use, as users cannot easily (or possibly even theoretically) follow why a system arrives at some output, which may undermine trust and make users less capable of responding to errors in the making \citep{mckinlay2020trust,kumar2024opacity}. In the military domain though, ``black boxes'' are especially problematic given how critical the chain of decision-making can be, both for purposes of contemporaneous oversight and correct judgment, and also for after-the-fact assessments of operations. Furthermore, black box systems are apt to entrench a degree of necessary unpredictability, as opaque systems can never be used with complete confidence in how they will behave in particular deployments, given that their workings are not understandable or transparent. This presents a special challenge for AI in the military, as uncertainty about a weapon's functioning can have a significant impact on the permissible uses of that weapon. More broadly, the impact of opacity on trust is particularly relevant, as humans employing an opaque AI system may have little reason to trust such a system, given that its workings are likely to remain a mystery, making it thus always at least partly unpredictable (from the human's perspective). 

Given the current methods for developing and training advanced AI systems (e.g., deep neural networks, machine learning, generative adversarial networks, etc.),\footnote{For a general overview of AI design choices, as well as some of the implications of such choices (with regards to characteristics such as opacity/transparency), see, e.g., \citep{duan2022survey,saghiri2022survey,speith2022review,cambria2023survey,li2023trustworthy}.} it is likely that many existing systems may never be capable of being transparent. Moreover, as systems become more advanced and are trained on larger amounts of data, transparency is liable to only become harder to achieve. In response to this, some have argued that we may ameliorate the problems posed by opacity by designing systems which are imbued with so-called ``explainable AI''. 

Roughly, ``[t]he purpose of an explainable AI (XAI) system is to make its behavior more intelligible to humans by providing explanations'' \citep[p. 1]{gunning2019xai}, and thereby to ``enable humans to understand, \emph{appropriately trust}, and effectively manage the emerging generation of artificially intelligent partners'' \citep[p. 83, emphasis added]{arrieta2020explainable}.\footnote{See also \citep{miller2019explanation,langer2021what,peters2022explainable} for ethical and social considerations regarding XAI. For surveys of XAI methodologies and findings, see \citep{adadi2018peeking,das2020opportunities,arrieta2020explainable,fiok2021explainable,speith2022review,cambria2023survey}. Cf. also \citep{ross2024ai} for criticism of the recent focus on transparency.} The underlying idea is that, in lieu of actual transparency -- which is likely to become increasingly unlikely, if not impossible -- we may get a functional equivalent of it through the creation of suitable explanations for an AI system's behavior. 

However, explainable AI models can be just as functionally opaque as systems without explanations \citep{rudin2019stop,zhang2019why,lakkaraju2020how}, and even when explanations might perhaps be of some use, they are apt to raise issues of their own with regards to human-machine teaming.\footnote{See, e.g., \citep{kwik2023performance,wood2024explainable} for discussion of challenges related to XAI in the military domain. For broader concerns with explanations' ability to improve responsible deployment of AI, see \citep{dorsch2024explainable,taylor2024explainable}.} More than this, assuming that explanations have value for actual deployers of AI-enabled systems, explanations will still only be useful so long as the system is trusted at a general level, and the explanations are trusted in particular. Thus, trust remains a central factor in not just human-machine teaming for opaque AI systems, but also for systems which have had opacity concerns ``addressed'' through XAI; unless we trust the system and its explanations, the explanations themselves aren't going to solve the problem. Furthermore, an AI which provides no explanations might sometimes be deemed more trustworthy (in terms of capability, reliability, ethics, and sincerity) than one which doesn't, because users may take the lack of explanations as a clear \emph{and honest} signifier that there are some things the human has to explore or judge for themselves. A system which provides explanations for everything, on the other hand, will at least in some cases present explanations which users judge to be wrong or misleading, and this is apt to greatly undermine trust, and do so in a way that is arguably more impactful than if explanations are simply absent from the beginning; a system which purports to do some thing but fails is arguably less trustworthy than one which never claims that capability at all. 

However, taken together, this should not be understood as an argument against XAI \emph{per se}. Indeed there is growing research showing how XAI may potentially be harnessed in various domains to mitigate some of the risks of opaque systems \citep{das2020opportunities,antoniadi2021current,machlev2022explainable,kalasampath2025review}. And as we will see in the following sections, added transparency in systems may provide opportunities for humans to more effectively and responsibly deploy AI-enabled systems. Moreover, XAI methods and tools, if thoughtfully developed and implemented, may help to mitigate some of the most pernicious risks associated with using AI-enabled systems. The points above highlight though that we should be cautious about technical solutions' ability to (fully) solve the problems presented by opacity. Rather, at the end of the day, in order for us to create effective and responsible human-machine teams where humans can trust AI-enabled systems, we must directly address the aspect of trust, and not hope to simply shoehorn it in through technical feats or design choices (even though those latter things may bring benefits of their own). 

The main contention of this section has been that trust does not require transparency, and that techniques solely aimed at mitigating the challenges posed by opacity (e.g., XAI) may sometimes be missing the mark. To drive this point home, we may conclude by considering some of the facets of trust discussed above. Consider a military AI system which is opaque, but which is tested across myriad conditions and always functions as intended, achieving the goals for which it was designed. It would seem clear that such a system would be deemed to be \emph{capable}, and thus trustworthy with respect to its capacity. Moreover, the more test conditions we pit the system against without it failing, the more we may deem it to be \emph{reliable}, and thus trustworthy with respect to this. If the system always opts for restraint in situations where targets are unclear or where collateral damage is expected to be high, we might also reasonably judge the system to be \emph{ethical} (even allowing for some awkwardness in conferring a trait like ``ethics'' to a machine). Thus, at least three of Malle and Ullman's four broad categories of trust would seem to be achievable for opaque systems. But beyond the human-machine paradigm, we may also consider that we trust other humans in every sense of trust explored by Malle and Ullman. Yet humans are fundamentally opaque. Whether or not we trust other individuals has nothing to do though with whether or not we have a clear view into their minds and can divine the processes which lead them to their decisions. Quite simply, trust does not demand transparency, and arguably never has. Rather, trust is built on something else, something which I will argue is far less precise, far harder to clearly demarcate, and far more \emph{human}, in a familiar, intuitive, and ``squishy'' way.

\section{Who's a good boy?}\label{sec_handlerapproach}

There are many ways we might think about trust and teaming, be it trust and teaming between peers, between supervisors and subordinates, between adults and children, or between humans and machines (or other man-made artifacts). 
Taking some of these more familiar sorts of trust and drawing out the dimensions explicated by Malle and Ullman, we might also hope, by analogy, to develop an understanding of human-AI trust which can be used to underpin strong human-machine teams. However, the simple problem is that machines are not people, nor will they behave like people or be trustable or trustworthy as people are. But as AI becomes more opaque, advanced, and potentially unpredictable, we cannot reasonably or responsibly continue to think of AI systems as ``simply machines'' either. Rather, AI systems will begin to occupy some sort of middle ground (if they don't already), where we are tempted to grant them increasing similarity to humans, in terms of how we trust them and team with them, but with us still retaining a sense that they cannot be treated exactly as humans would be. The question, at that point, is how we should best treat them and treat with them? 

In the literature on AI in the military, especially AI which is opaque and may act in potentially unpredictable ways, there have been many analogies used to grapple with the challenges of teaming humans with such systems, of locating responsibility for the outcomes brought about by such systems, or even simply for our general understanding of how we should best understand these systems within the structures of military hierarchies and command and control networks. Some authors have discussed opaque and unpredictable systems using the language of child soldiers (\citep[pp. 73--75]{sparrow2007killer}; \citep[ch. 9]{galliott2015military}; \citep[pp. 73--76]{crootof2018analogy}). However, while this analogy gets some things right, it unduly taints the discussion in virtue of the many and significant moral and legal issues with employing child soldiers. Moreover, given how unpredictable children may be, especially children subjected to the stresses and terrors of war, the analogy sets a baseline of unpredictability which is inapt for AI-enabled and autonomous systems used in the military, even ones which are opaque and \emph{potentially} unpredictable.\footnote{It is worth noting that we assume throughout that systems potentially deployed are, at a minimum, designed in accordance with the demands of international humanitarian law. This means that such systems must be able to be used in a discriminate manner, which requires that they be reliable and predictable enough that combatants can control the effects of those systems. All systems will admit some level of unpredictability, even wholly deterministic systems, but if some autonomous or AI-enabled system were to be fundamentally or egregiously unpredictable, its use would be impermissible.}

A more promising analogy increasingly used in military science circles is to view military AI through the lens of centaurs, mythical creatures that are half-man, half-horse \citep{macak2005centaurs,scharre2016centaur,neads2021tools,johnson2023ai}. This understanding has proven highly effective for assessing and improving human-machine teaming. However, by looking to an analogy which centers around the \emph{fusion} of man and beast, the centaur model is in some respects lacking. Most importantly, it glosses over the ways in which autonomous and opaque AI-enabled systems may execute processes independently of humans, rather than as extensions of direct and contemporaneous human intent. This is critical, as many of the deepest concerns about military AI and autonomous weapon systems center around the dangers posed by systems bringing about outcomes without human oversight and indeed against human intent. The analogy also shifts how we might view trust within human-machine teams, as the ways we trust an extension of ourselves are not the same as how we may trust other agents or entities with whom we are collaboratively working toward some goal. Furthermore, by looking at a fusion-oriented understanding, the centaur analogy and model clouds the fact that for opaque and potentially autonomous AI systems, the role of humans teamed with them will often be more about managing the systems and broadly directing them toward proper outcomes than about humans directly having control over the systems in real-time. In short, the centaur analogy and model, while providing much benefit to the debates on human-machine teaming, still fails to fully account for both the capacity of AI systems to independently execute processes and for our attendant responsibility to handle these systems with judgment and care.\footnote{Two underlying points which are critical to these arguments are that 1) machines do not act -- in the philosophical, ethical, or legal sense of ``acting'' -- but only ever execute processes, and 2) machines can be designed and deployed so as to extend or offload a human's will, and which of these is pursued (at the design and deployment stage) will have significant bearing on further questions of responsibility, teaming, and other concerns. See [author's citation removed]; [author's citation removed] for exploration of these points.} In order to remedy this, we must look to an analogy which highlights the impact we can have on the behavior of AI systems, but which bears constant vigilance to their independence and potential unpredictability. Enter Buddy, the Labrador Retriever.


\subsection{Buddy the Dog}\label{subsec_dogs}

As any pet owner knows, your dog or cat will display a variety of behaviors, most of which become fairly predictable over time and as you become more and more familiar with the pet. Moreover, while a pet is obviously an independent entity capable of doing as it wishes, even against your commands,\footnote{In such cases, a guilty expression might be predictable. Bad dogs are often insufferably cute dogs.} it is also clear that most animals are capable of being trained to reliably do a wide variety of tasks, many of them quite complex and requiring at least some degree of planning \citep{waller1958dogs,anderson2001teams,hammerstrom2005ground,briefer2024goats,lee2024why}. Further still, animals have long been used in war, and still are, and animal combatants are granted a fair degree of trust from their human counterparts, despite the fact that animals are opaque and may act independently of and even against the wishes of their human handlers \citep{waller1958dogs,cruse2014military}. Given these factors -- that animals may be rigorously trained, highly reliable, tightly connected to humans overseeing them, but still capable of acting unpredictably and in unwanted ways -- the analogy between animals and autonomous or AI-enabled systems provides us with a strong starting point for better exploring trust and teaming between humans and machines.\footnote{For similar (but brief) arguments, see \citep[pp. 10--12]{wood2023awsresponsibility}; \citep[pp. 9--11]{wood2024explainable}. See also \citep{sullins2011robot,roff2018trust,crootof2018analogy,baker2022should} for similar analogical discussions. While the animal-robot/AI analogy is not new, most accounts utilize it to shed light on how to best understand robots/AI or aspects of agency. The account developed here, however, focuses on how the analogy can be leveraged to help us better understand and develop human-machine teams. Thus, while earlier works focused more on the ontological or metaphysical questions of what AI is or whether it should be considered an agent, this work's emphasis is on practical upshots of the analogy and how it can be utilized to improve real-world human-machine teams.}

Looking more closely at the analogy, we can see that there are numerous and critical touch-points between animals and AI systems. Both animals and AI systems will be endowed with certain ``hardware'' which sets limits on what may be achievable or which instigates in favor of certain capabilities, courses of action, or general approaches to problem-solving. For example, a Labrador Retriever and a crow are both highly intelligent and may both be trained and given tasks by a human trainer/handler. However, mere intelligence and trainability does not mean both may be given the same tasks, or, if given the same tasks, that both will be equally successful. Rather, their physical makeup fundamentally impacts on what these creatures will be able to do, just as the hardware of a given AI system will shape what it may do. 

On a related point, just as an AI may be given certain software to help guide its behavior, we may also think of an animal's training in a similar light; in virtue of teaching and entrenching certain visual cues, verbal commands, or other means of communication, two animals which are otherwise the same physically may be given widely divergent response patters to one and the same stimulus. The way(s) an animal is taught can also alter its general approach to problem-solving or its mode of carrying out tasks. Consider, for example, a Labrador trained as a hunting retriever and one trained as a service animal for someone who is visually impaired. If both are told to ``fetch'' some object, both are likely to do so successfully, but the former will bolt as fast as it can and often with little regard for what obstacles are in its way, whereas the other will take more care and ensure that, first and foremost, its charge is not put into danger. One and the same task executed well in either instance, but executed in vastly different ways, simply due to training. And the same holds for AI systems as well. 

More importantly than hard- or software though is that for both animals and man-made systems, trust is often far more about familiarity than about transparency, a deep understanding of a system, or any significant level of technical knowledge. This is evidenced by the fact that owners of a pet may often have much more trust in their animals than an animal psychologist might, despite the fact that the latter generally will have far more knowledge about why an animal does what it does than its owners would.\footnote{Trust borne of familiarity may not translate to predictability directly though. Rather, at least some degree of technical knowledge will be necessary. Yet even so, owners who are familiar with their pets, though not always as successful in predicting behavior, still have an understanding for when things are ``off'' with their animals. See respectively \citep{demirbas2016adults} and \citep{demirtas2023dog}.} The problem is that knowledge does not necessarily equate to understanding, in the squishy, hard-to-define manner in which two people might ``understand one another''. The same holds for users of manmade systems in comparison to designers, engineers, or technicians of those same systems. From pilots ``trusting'' their aircraft to tank commanders ``feeling that something is off'' in their vehicle, humans who regularly rely on machines often develop subtle and insightful understandings of those systems. Moreover, the understanding had by a regular user of a vehicle, weapon system, program, or even AI will be far different than that had by someone who has knowledge of that system but lacks familiarity. 

Beyond this, familiarity not only contributes to trust, but also to predictability. This is because humans can take in many pieces of information from our environment and process them in subtle ways which are not necessarily readily apparent to us. This is, indeed, what allows pilots, tank commanders, programmers, or any other regular user of a system to have that unexplainable understanding of when things will go awry. It is not necessarily that some precise stimulus can be pointed to as \emph{the} indicator that there will be a problem, but knowledge had through intimacy and regular use allows for predictions even when the predicting individual does not know why they are so certain that things will transpire a particular way. If some degree of technical knowledge is added, this only strengthens one's ability to understand what is going on in a system.\footnote{This is one area where, \textit{pace} the arguments of Section \ref{sec_transparency} above, explanations and XAI can prove useful. \textit{If} a handler possesses the necessary familiarity with a system, then providing explanations could help bolster predictability or effective and responsible use. It is also worth noting though that explanations could also create a false sense of security, dulling the impact of a handler's intuitions regarding how an AI-enabled system might behave in some scenarios. Thus, explanations may cut both ways, and only thorough empirical research into actual human-machine teaming (with and without XAI) can fully address this concern. Thanks to an anonymous reviewer for pressing me to more carefully consider how XAI may prove useful and what the limits of that use may be.}

The same holds for animals as well, and the central factor in human-animal teams is not that the animal is wholly transparent or always clearly predictable -- they clearly aren't -- but rather that the handler has a ``feel'' for the animal and can ``sense'' when conditions may make it liable to misbehave. That ``feel'' is just as important for human-machine teams.

Finally, it is worth stating clearly that human-animal teams are founded on a particular type of relation, namely that between a handler and a handlee. Though it may not seem self-evident, the same is true for humans ``using'' or ``controlling'' opaque and potentially unpredictable AI systems as well. Insofar as humans cannot really know for certain what these systems will do, we cannot see ourselves as ``using'' them in the same way we would a simple(r) and fully mechanistic/deterministic tool. Rather, we must recognize that such AI systems represent somewhat agent-like entities which we are handling and which may ``go haywire'' for reasons we don't fully understand. 

By recasting our view of the relationship between humans and AI systems, from the user-system paradigm to a handler-handlee paradigm, we focus on this fact, that opaque systems raise the same worries brought forward when we deploy animal combatants to war zones. The problem is not that these may sometimes act unpredictably, but rather that we refuse to use the language and reasoning we would apply to other entities which are relevantly similar. By thinking in terms of humans-as-handlers of AI systems, we re-anchor responsibility for the outcomes brought about by such systems and make clear that humans tasked with handling or overseeing them must have not just the technical expertise, but also the familiarity and attendant trust needed to know when it is ethical and legal to ``take off the leash''.\footnote{For brief arguments along similar lines, see \citep[pp. 10--12]{wood2023awsresponsibility}; \citep[pp. 9--11]{wood2024explainable}.} 

\subsection{RoboBuddy}\label{subsec_robots}

If we are to take the analogy of human-animal teams as our starting point for thinking about human-AI teams, it is important that we not just highlight the similarities, but also pay heed to the stark differences between animals and machines. In particular, we must be mindful of the \emph{dis}analogies which require us to rethink certain aspects of the humans-as-handlers approach as inspired by human-animal teaming. 

First, unlike animals used in war, military systems imbued with AI may be networked, allowing for mistakes or failures within a single AI system to be propagated across large numbers of combat platforms. This makes it far more important that such systems be tied very closely to human handlers' actions in ways that go far beyond what is demanded in human-animal teams. This is for the simple reason that an AI distributed across many combat platforms can be far more destructive than a single animal combatant when both are ``taken off the leash'', so to speak. The ways that animals and AI systems may fail also greatly differ. Where an animal may make mistakes due to fear, frenzy, or other emotional/psychological stimuli which alter judgment, AI systems will not be subject to these factors. Rather, AI systems may make mistakes due to things like programming errors, electronic warfare meant to confuse or disrupt the AI system, or training failures related to bad data sets, or, worse still, poisoned data. This also points to a further factor regarding errors/mistakes, namely that for animals, mistakes are liable to be more tightly connected to very discrete contexts and sequences of events, whereas for AI systems mistakes may often be indicative of larger failings in the system in dire need of addressing. The divergence in how and why things may go wrong for animals versus AI systems further affects how humans must be teamed with both, in particular with what cues human handlers must look for when determining whether it would be safe and responsible to allow an animal/AI to make lethal decisions in some (combat) environment. 

In addition to core differences between the entities themselves -- animals vs. AIs -- there are also liable to be stark differences in how we view and relate to both. Humans are apt to anthropomorphize machines, especially AI-enabled ones which exhibit behaviors that appear spontaneous and indicative of some sort of agency \citep{salles2020anthropomorphism,johnson2024finding}, and humans whose safety depends on a machine, even a simple(r) one such as a manned aircraft or ground vehicle, may lean into some degree of anthropomorphization \citep{simmons2010humanizing,anderson2020bomber}. However, such extension of agency is unlikely to amount to that which is conferred to animals, especially animals working in tight human-animal teams. Furthermore, the dimensions of trust shared between a human handler and their animal charge are apt to be much more intimate and involve more moral aspects of trust than could be expected within human-machine teams. This requires us to be careful about what sorts of trust we can expect to translate from the analogy of human-animal teams to the case of human-machine or human-AI teams. 

Beyond this, we may also wish to actively discourage certain forms of trust and bonding germane to human-animal teams from being ported into human-machine teams. Considering that machines designed for warfighting ought ideally to be attritable, are likely to be replaceable by functionally identical replenishments, and are intended to offset some of the risks currently imposed on human combatants, if humans become too attached to their AI-enabled handlees -- as they likely would be with say, combat assault dogs they are teamed with -- then this may undermine some of the very reason we have for using such systems.\footnote{For such arguments, see, e.g., \citep{cappuccio2021saving,mamak2023military}.} Indeed, humans have already shown deep connections to robotic systems, and ones which are sometimes much simpler than the current and near-future autonomous and AI-enabled platforms we are likely to see; over 15 years ago, one could already find stories where ``[o]ne EOD [explosives ordnance disposal] soldier brought in a robot for repairs with tears in his eyes and asked the repair shop if it could put `Scooby-Doo' back together'', or ``another EOD soldier who ran 164 feet under machine gun fire to retrieve a robot that had been knocked out of action'' \citep{hsu2009real}. If bonding and trust lead soldiers to place more value on machines than on a human's life or safety (their own included), this would undermine much of the reason for developing and deploying such systems in the first place. Moreover, the fact that machines can be replaced must be anchored as a central tenet in training humans for human-machine teams. Critically, we must bear in mind that finding the right balance between a high enough bond and trust for effective and responsible teaming will generally cut against a users' ability to emotionally and psychologically detach from that system. Finding that balance will thus be a recurrent institutional and training challenge. It will also likely have no final ``solution'', presenting new and different obstacles with each generation, generations which generally have subtly different emotional, psychological, and moral sentiments when compared to their predecessors. Thus, we must be careful to establish human-machine teams in such a way that we gain the benefits, in terms of trust, effectiveness, and responsibility, of strong human-animal teams without carrying over the additional emotional attachments which are almost guaranteed to be included in human-animal teamings, and we must recognize that the balancing act of doing so will be a challenge of enduring relevance. 

\section{Humans-as-Handlers: A collaborative approach}\label{sec_humansashandlers}

If we are to model human-AI teams after the fashion of human-animal teams, how do we take the fundamentals from the latter and build them up within the former, and once that is achieved, how do we cull any unwanted attitudes or sentiments which may attend human-animal teams? In this section, we will look at the entire life-cycle of an AI system, from design to training and up through deployment, exploring a formulation of the humans-as-handlers approach which is collaborative and puts front and center the importance of opaque AI systems' potential unpredictability and capacity to act independently of and potentially against the intentions of human handlers. It is also worth stating clearly that the purpose of this section is merely to provide a sketch of the humans-as-handlers approach to human-machine teaming for opaque AI systems; a full accounting of the technical, human, and institutional challenges involved in pursuing this approach, as well as how we can meet those challenges, must be left to further work involving not just theoreticians, but also policy-makers and military professionals. For concreteness in the following, the discussions of human-animal teams will look explicitly to the template provided by humans teamed with dogs, in particular combat assault dogs used in war. 

To begin with, the \emph{design and development} of an AI program or system represents the first stage in its life-cycle. At a general level, how a system is designed, what methods are used and what architectures chosen, will not necessarily have a significant impact on the precise shape of the human-machine teams which may later be crafted to harness that AI system's potential. However, certain design choices may well make certain tasks more achievable or create opportunities (or challenges) for specific efforts at human-machine teaming. For example, though it was argued above that transparency and explainability will not alone suffice to create trust or underpin strong human-machine teams on their own, these factors can contribute to better teaming by opening up additional possibilities for human handlers to better understand their artificial charges \citep{wood2024explainable}. Alongside familiarity, a system which provides information which is understandable to a handler could plausibly be more reliably and responsibly deployed, in virtue of having auditable processes which can be interrogated for potential mistakes lying in wait. That being said, it is important to be clear that handlers do not need to know in a mechanistic, deterministic, or transparent way why a system does what it does, just as a handler of a combat assault dog does not need to know these things in this manner. What counts is that handlers have an intuitive understanding of the machine's/dog's behavior and can reliably predict what situations are apt to make the machine/dog make mistakes. In short, the key is not transparency, but rather familiarity. 

This leads to the second stage of an AI system's life-cycle, namely the \emph{training} phase. Though often overlooked in the literature on human-machine teaming, this is a critical point at which to begin teaming eventual handlers with AI systems. This is because handlers will only have an intuitive understanding of when, why, and how their AI systems may fail if they are familiar with the ways AI systems are trained, with the data they are trained on, with the sorts of mistakes the AIs make during training, and with how those mistakes may be addressed through remedial training regimen. In short, unless handlers know what it is that makes their AI charges ``think'' (for lack of a better word) as they do, those handlers will not know when the AI is liable to err. Building on this, handlers involved in the training process should be encouraged to actively attempt to trick or ``break'' the AI system; only by knowing firsthand its limits, how it reaches those, and what may be done to counteract its failings can handlers really have confidence in how reliably they can control their AI systems.\footnote{\label{ftnote_woodreliability}This touches on the fact that ``reliability'' is a highly context- and user-dependent factor. See, e.g., \citep{wood2024reliability}.} In attempting to outsmart their AI charges, handlers will also gain an appreciation for the fact that all training is contextual, and only through a sufficiently broad set of training environments can an AI system be expected to, with any reliability, function well. Moreover, this would highlight for handlers that novel environments will always prove challenging for AI systems, a fact that must be born centrally in mind when considering the use of these in critical domains like warfighting. 

By forcing attention on the training phase, the human-animal analogy also highlights a recurrent failing in ongoing debates; much of the discourse on human-machine teaming centers around how to embed and incorporate AI systems into military hierarchies or missions, but this often implicitly assumes that such AIs are already trained. This is a problem, as the combatants tasked with handling AI systems must have a familiarity with not just the AI as an end-product, but also as a system in training, as the training regimen itself includes significant lessons on how, when, why, and to what severity the AI may make mistakes. In the same way that you cannot simply buy a trained dog and fully expect it to listen to you (because you lack authority and familiarity with it), you cannot simply buy an AI and expect to be able to completely effectively and responsibly use it, for the simple reason that there will be many nuances to how it responds to stimuli, and the mere purchaser who sees himself as a ``user'' of the system will not know what cues to look for in order to know when the system may be about to act in an unpredictable way. In short, just as handlers of combat assault dogs work with those dogs throughout their training to develop a bi-directional trust which underpins their collaborative efforts, human-machine teams involving opaque AI systems must likewise involve human handlers throughout the training process. Only by doing so can those handlers have the necessary understanding and intuition needed to responsibly and effectively harness the potential of AI systems in war. 

Once a system has been trained, ideally with the input and presence of the eventual handlers of that system, the AI will then be deployed to actual environments where it is expected to carry out tasks. At this \emph{deployment} stage, the humans-as-handlers approach adds value in multiple ways. First, by clarifying who is responsible for the outcomes brought about by the AI system, it helps address the wealth of concerns of so-called ``responsibility gaps''; just as a handler of a combat assault dog bears responsibility for mistakes the dog may make (after all, accounting for such mistakes when deciding to take off the muzzle is a core job of the handler), an AI's handler is responsible for what the AI does.\footnote{For discussion of and response to the responsibility gap, especially in the military, see, e.g., \citep{matthias2004responsibility,asaro2006should,sparrow2007killer,zajac2021punishing,koenigs2022artificial,oimann2023responsibility,wood2023awsresponsibility}.} Related to this, showcasing the responsibility had by a clear set of individuals -- i.e., handlers of AI systems -- highlights lines of command and control as well; officers may give orders to AI handlers to, with the aid of their AI-enabled combat systems, carry out certain tasks, but it is incumbent upon those handlers to evaluate whether it would be permissible under the laws of war to actually employ their AI-enabled systems when in the field. If handlers deem that it would be too dangerous, risk too much harm to noncombatants, etc., then it is their job to halt a strike using the AI system. Thus, we not only trace responsibility for bad outcomes, but also reduce the likelihood of bad outcomes occurring, by delineating a clear set of individuals whose responsibility it explicitly is to avoid such outcomes. And if handlers have been involved throughout the training process, as argued for under the humans-as-handlers approach, they will be likely to have a fair grasp of what sorts of things may trip up the system, will have a better idea of how to mitigate risks posed by contexts which have previously been challenging for the AI, and, most importantly, they will have firsthand experience of those cases where the AI will expectably be unpredictable, or will predictably misbehave.\footnote{Again, predictability can be user-dependent. See note \ref{ftnote_woodreliability} above.} Individuals who are genuinely familiar with a system (organic or man-made) will also know that you cannot be sure how it will behave in novel environments. Finally, the familiarity demanded and provided by the approach means that handlers will in general have more reliable ``gut feelings'' about their AI charges, providing them with a faster check on the reliability of a system in discrete deployment contexts, an important factor in warfare where speed can be of the essence and there may be little time to double-check one's data or explanations. In such environments, it becomes increasingly crucial that ``gut feelings'' be well-honed and based on trustworthy foundations (like extensive past experience collaborating in the training of a system), and the humans-as-handlers approach argues for just that. The approach thus entrenches safety and responsibility at numerous levels, all while retaining and arguably enhancing the effectiveness of AI systems used in war. 

Post-deployment, there will be a \emph{troubleshooting} phase during which after-action reports are drafted, diagnostics and troubleshooting are carried out for all assets used (including AI-enabled assets), and repairs and improvements are made where necessary. Just as handlers of combat assault dogs accompany their canine comrades in virtually all phases and aspects of a deployment \citep{cruse2014military}, handlers of AI systems should also accompany these systems throughout the post-deployment phase. This is because handlers, due to their expertise and deep familiarity/understanding of their AI systems, will likely be in a strong position to help technicians more quickly find and address any problems that were encountered in the field. This has clear military benefits, in terms of speeding the refurbishment process for combat assets. More importantly for trust and teaming though, handlers should be involved in any troubleshooting and repair works for their AI charges, as it is important for handlers to know why things may have gone wrong, and to know what is being done to their AI comrades to address the situation. Furthermore, there are reasonable grounds to include handlers in any decisions regarding the retraining and/or reprogramming of an AI system, should it make mistakes in the field which lead to unwanted results; as the individuals responsible for the AI systems and the ones who are expected to carry responsibility for mistakes, a full measure of respect for the handler-handlee relationship naturally implies that handlers would have some broad say over what happens when things go awry. Beyond this, a feeling of trust that extends beyond the battlefield is apt to create a sense of enduring camaraderie that leads to stronger teaming overall, and to trust which goes beyond the mere \emph{performance trust} usually associated with man-made artifacts.

Overall, though the human-animal team sets a strong foundation for human-AI teams, we must also be sure that handlers understand that AI-enabled systems are not agents (at least, not yet). We must thus ensure that human handlers are not placing undue moral or other weight on the value of their AI charges. This creates a tension, as the establishment of strong teams rooted in trust and familiarity is apt to create deep connections between handlers and the systems they are overseeing. However, this challenge may arguably be addressed through socio-technical solutions, such as teaming humans with particular AI programs and then embodying the AI in combat platforms prior to deployment. In this manner, the loss of the combat platform would not constitute the loss of the AI itself, allowing the human handler to be properly familiar with and responsive to the AI system operating the platform while still being capable of treating the platform (but not the AI) as an attritable asset. Exactly how to best balance bonding and teaming against an ability to sacrifice systems when necessary will be an ongoing institutional and training challenge though, and one deeply connected to the current psychological and emotional norms of a given generation of handlers. On a related note, though the humans-as-handlers approach takes as a departure point the relationship between human handlers and dogs used in war, looking to the bonds of trust and teaming found in this partnership, there will be important aspects of the human-animal team which likely cannot be included in human-machine teams. In particular, the sincerity dimension of trust is apt to be a critical difference; dogs and other animals can demonstrate sincerity to humans, and vice versa, but it is unclear how an AI even could be sincere. True, AI systems can likely be trained to recognize cues for sincerity from humans, and can be programmed to provide cues in return, but if these are viewed as simple programming constructs and not as genuine displays of sincerity (a plausible assumption, at least for some human handlers' perceptions), then this may undermine trust. In any case, sincerity can be an important grounding dimension of trust from which other dimensions are given further stability and weight, and it represents an important difference between AIs and animals which must be borne in mind when actualizing the humans-as-handlers approach. 

Finally, no matter how many dimensions of trust may plausibly be captured for AI systems, humans and AI systems must act \emph{collaboratively}.\footnote{In the military domain, this point is also echoed by \citep{macak2005centaurs,neads2021tools,trevithick2024everything,tsamados2024human}. See respectively \citep{nyholm2017attributing} and \citep{gundersen2022future} for similar remarks regarding AI in general and within medicine. For a critical view, see \citep{evans2023collaborate}.} Centrally, this means that AI systems cannot be viewed as tools that are ``used'', but rather as teammates with whom one collectively pursues goals. And whether or not AIs are truly agents does not matter to this point. The central tenet is not that AI systems \emph{are} agents with whom handlers collaborate, but rather that handlers \emph{should view them as such}. Thus, it is about the handler's perception, and not about an ontological fact concerning the AI systems. This is important to make clear, as the opacity and potential unpredictability of AI systems makes it risky (if not negligent) to view these as tools that can be ``used'' as one would a hammer. But by viewing AI systems as collaborating partners akin to dogs, we elevate these systems in terms of their perceived autonomy and potential unpredictability, while still relegating them to a status which is clearly below the level of a human agent. This provides dual benefits by ensuring that handlers will give AI systems due respect and care, while also making clear that it is handlers who are in charge, who call the shots, and who ultimately bear responsibility for the outcomes brought about by the human-machine team. 

\section{Conclusion}\label{sec_conclusion}

Though actual efforts at human-machine teaming vary across states and organizations, there is a broad recognition that such teaming is necessary, especially in high-stakes environments like warfighting. In the United States, both Army and Air Force leaders have spoken to the critical necessity of developing and entrenching strong human-machine teams \citep{seffers2023man,park2024mastering}, and while it does not provide a guide to such teaming, the United States Department of Defense's (DoD) recent \textit{Directive 3000.09: Autonomy in Weapon Systems} provides many statements which fit well with the humans-as-handlers approach advocated for here. There is emphasis on personnel being able ``to exercise appropriate levels of human judgment over the use of force'', all ``while remaining responsible for the development, deployment, and use of AI capabilities'' \citep[pp. 3, 6]{usdod2023directive3000.09}. The DoD also states that ``AI capabilities will be developed and deployed such that relevant personnel possess an appropriate understanding of the technology, development processes, and operational methods applicable to AI capabilities'' \citep[p. 6]{usdod2023directive3000.09}. Exactly how such ``understanding'' is to be established is, however, not explored. The humans-as-handlers approach details one way for doing precisely this, and alongside the development and deployment guidelines of documents like \textit{Directive 3000.09}, it provides a means for harnessing not just technological approaches to human-machine teaming, but also institutional and human approaches. 

In order for us to truly trust the emerging generation of autonomous and AI-enabled systems, especially as these are increasingly used in domains such as warfighting, we will need more than mere explanations or transparency. Rather, we will need humans to be intimately and habitually teamed with the AI-enabled systems they are intended to oversee. In order to properly effectuate such teams, we must also dispense with the language of humans as ``users'' of opaque AI systems and instead recognize that humans will be ``handlers'' of such systems, in many ways akin to those humans who work as handlers of combat assault dogs, bomb defusing dolphins, or other animal combatants already incorporated into military structures. The necessity for bonds of trust built on intimacy and familiarity is further underscored by the increasing speed of modern conflict, where perusal of explanations or oversight of a system's decision processes is apt to be too unwieldy to allow for meaningful control in the field; we must have handlers who can ``feel'' that ``something is off'' at an intuitive level and then act to intervene before it is too late. Anything less allows for the risks of opaque and potentially unpredictable AI systems to remain unchecked, opening the military to serious ethical and legal challenges. 

In future research, it will also be useful to examine to what extent modeling human-AI teams after human-animal teams may affect broader public trust in the implementation of AI systems within domains like warfighting or medicine. This is important, as much of the literature on trust and teaming for AI systems is focused on these elements \emph{within} human-machine teams, but there is far less attention paid to how ways to improve intra-team trust may impact on the overall trust society may be willing to give to human-AI teams. Given that skepticism of AI within expert communities and skepticism in the lay public may be rooted in different things, we must be mindful of the fact that for AI to truly present benefits to society, the humans handling such systems must have enough trust to actually use the systems effectively, but society itself must also trust the human-AI teams enough to grant them permission to oversee critical domains like state security or public health. It seems plausible that modeling human-machine teams after human-animal teams, a relatively familiar and innocuous relationship understood by every pet owner, would present a strong route for entrenching such broader public trust. This remains to be shown in further research though. 

On a related point, there may be worries that over-reliance on an analogy to animals, especially familiar animals like dogs, may reduce scrutiny of AI-enabled systems or engender inappropriate and potentially unwarranted levels of trust in these. At the level of a whole society, if AI systems begin to share an intuitive intellectual space with animals, this is also sure to have trickle-down effects on handlers of discrete systems tasked with discrete missions or goals. These are interesting and important questions, but ones which will require involved psychological research spanning across time and across regions and domains of potential AI deployment. It is critical that such research be carried out going forward though, as what ethical and legal challenges may arise will be highly contingent on how such systems are received by the broader public, and speculation at this moment is liable to provide, at most, a foggy notion of the issues lying in wait.\footnote{My thanks to an anonymous reviewer for pressing me to consider these broader possibilities.} It is also worth stating that while we have focused on familiar human-animal teams, especially between humans and dogs, there is value in exploring other human-animal teams as well. Drawing out differences between certain animals in how they are teamed with humans may provide valuable insight into how we might best approach human-machine teaming for divergent autonomous and AI-enabled systems.

At any rate, it is clear that while the myriad of technical achievements in explainability and transparency do provide some benefit for human-AI teaming, in order to responsibly and effectively employ opaque AIs in sensitive domains, we will need a thick and reliable form of trust. Moreover, that trust must be built not on academic understandings of how reliable an AI is for some task, but rather on an intimate sense of an AI system's reliability across various contexts, on an understanding born of familiarity and having a ``feel'' for a given system. Given the need for speed in critical endeavors such as medicine or the military, anything less than this will leave handlers of such AIs without a strong guide for judging when things may go awry, leaving them incapable of doing one of the most central jobs of a handler, namely intervening to stop tragic mistakes before they happen. 

\pagebreak 
{\small \bibliography{master}}								

@Article{roff2018trust,
  author    = {Roff, Heather M. and Danks, David},
  journal   = {Journal of Military Ethics},
  title     = {``{T}rust but Verify'': The Difficulty of Trusting Autonomous Weapons Systems},
  year      = {2018},
  number    = {1},
  pages     = {2--20},
  volume    = {17},
  publisher = {Taylor \& Francis},
}

@Article{zajac2021punishing,
  author    = {Zaj{\k{a}}c, Maciej},
  journal   = {Journal of Military Ethics},
  title     = {Punishing Robots -- Way out of {S}parrow's Responsibility Attribution Problem},
  year      = {2021},
  number    = {4},
  pages     = {285--291},
  volume    = {19},
  doi       = {10.1080/15027570.2020.1865455},
  publisher = {Taylor \& Francis},
}

@Book{baker2022should,
  author    = {Baker, Deane},
  publisher = {Polity},
  title     = {Should We Ban Killer Robots?},
  year      = {2022},
  isbn      = {9781509548514},
  series    = {Political Theory Today},
  url       = {https://books.google.de/books?id=hw6BzgEACAAJ},
}

@TechReport{hrw2018heed,
  author      = {{Human Rights Watch}},
  institution = {Human Rights Watch},
  title       = {Heed the Call: A Moral and Legal Imperative to Ban Killer Robots},
  year        = {2018},
  publisher   = {Human Rights Watch},
  url         = {https://www.hrw.org/sites/default/files/report_pdf/arms0818_web.pdf},
}

@Article{chengeta2016accountability,
  author    = {Chengeta, Thompson},
  journal   = {Denver Journal of International Law \& Policy},
  title     = {Accountability Gap: Autonomous Weapon Systems and Modes of Responsibility in International Law},
  year      = {2016},
  pages     = {1--50},
  volume    = {45},
  publisher = {Hein Online},
}

@TechReport{hrw2012losing,
  author      = {{Human Rights Watch}},
  institution = {Human Rights Watch},
  title       = {Losing Humanity: The Case Against Killer Robots},
  year        = {2012},
  publisher   = {Human Rights Watch},
  url         = {https://www.hrw.org/sites/default/files/reports/arms1112_ForUpload.pdf},
}

@Article{asaro2006should,
  author  = {Asaro, Peter M.},
  journal = {International Review of Information Ethics},
  title   = {What Should We Want From a Robot Ethic?},
  year    = {2006},
  pages   = {9--16},
  volume  = {6},
}

@Article{matthias2004responsibility,
  author    = {Matthias, Andreas},
  journal   = {Ethics and Information Technology},
  title     = {The Responsibility Gap: Ascribing responsibility for the actions of learning automata},
  year      = {2004},
  number    = {3},
  pages     = {175--183},
  volume    = {6},
  publisher = {Springer},
}

@Article{pagallo2011robots,
  author    = {Pagallo, Ugo},
  journal   = {Philosophy \& Technology},
  title     = {Robots of Just War: A Legal Perspective},
  year      = {2011},
  number    = {3},
  pages     = {307--323},
  volume    = {24},
  doi       = {10.1007/s13347-011-0024-9},
  publisher = {Springer},
}

@Article{sharkey2007automated,
  author    = {Sharkey, Noel},
  journal   = {Computer},
  title     = {Automated Killers and the Computing Profession},
  year      = {2007},
  number    = {11},
  pages     = {124--123},
  volume    = {40},
  publisher = {IEEE},
}

@Article{crootof2016war,
  author    = {Crootof, Rebecca},
  journal   = {University of Pennsylvania Law Review},
  title     = {War Torts: Accountability for autonomous weapons},
  year      = {2016},
  pages     = {1347--1402},
  volume    = {164},
  publisher = {Hein Online},
}

@Article{wagner2014dehumanization,
  author    = {Wagner, Markus},
  journal   = {Vanderbilt Journal of Transnational Law},
  title     = {The Dehumanization of International Humanitarian Law: Legal, ethical, and political implications of autonomous weapon systems},
  year      = {2014},
  pages     = {1371},
  volume    = {47},
  publisher = {Hein Online},
}

@Article{crootof2018analogy,
  author    = {Crootof, Rebecca},
  journal   = {Harvard National Security Journal},
  title     = {Autonomous Weapon Systems and the Limits of Analogy},
  year      = {2018},
  pages     = {51--83},
  volume    = {9},
  publisher = {Hein Online},
}

@Article{koenigs2022artificial,
  author    = {Peter K\"{o}nigs},
  journal   = {Ethics and Information Technology},
  title     = {Artificial Intelligence and Responsibility Gaps: What is the problem?},
  year      = {2022},
  number    = {3},
  volume    = {24},
  doi       = {10.1007/s10676-022-09643-0},
  publisher = {Springer Science and Business Media {LLC}},
}

@Article{oimann2023responsibility,
  author    = {Oimann, Ann-Katrien},
  journal   = {Philosophy \& Technology},
  title     = {The Responsibility Gap and {LAWS}: A Critical Mapping of the Debate},
  year      = {2023},
  number    = {3},
  pages     = {1--22},
  volume    = {36},
  doi       = {https://doi.org/10.1007/s13347-022-00602-7},
  publisher = {Springer},
}

@Article{wood2023awsresponsibility,
  author    = {Wood, Nathan Gabriel},
  journal   = {Ethics and Information Technology},
  title     = {Autonomous Weapon Systems and Responsibility Gaps: A taxonomy},
  year      = {2023},
  number    = {1},
  pages     = {1--14},
  volume    = {25},
  doi       = {10.1007/s10676-023-09690-1},
  publisher = {Springer},
}

@TechReport{usdod2023directive3000.09,
  author      = {{US Department of Defense}},
  institution = {United States Department of Defense},
  title       = {{DoD Directive} 3000.09},
  year        = {2023},
}

@Article{fiok2021explainable,
  author    = {Fiok, Krzysztof and Farahani,Farzad V. and Karwowski, Waldemar and Ahram, Tareq},
  journal   = {The Journal of Defense Modeling and Simulation: Applications, Methodology, Technology},
  title     = {Explainable artificial intelligence for education and training},
  year      = {2021},
  number    = {2},
  pages     = {133--144},
  volume    = {19},
  doi       = {10.1177/15485129211028651},
  publisher = {{SAGE}},
}

@TechReport{michel2020black,
  author      = {Michel, Arthur Holland},
  institution = {{U}nited {N}ations Institute for Disarmament Research ({UNIDIR})},
  title       = {The Black Box Unlocked: Predictability and Understandability in Military {AI}},
  year        = {2020},
}

@Article{das2020opportunities,
  author    = {Das, Arun and Rad, Paul},
  title     = {Opportunities and Challenges in Explainable Artificial Intelligence ({XAI}): A Survey},
  year      = {2020},
  doi       = {10.48550/ARXIV.2006.11371},
  publisher = {arXiv},
}

@Article{gunning2019xai,
  author    = {Gunning, David and Stefik, Mark and Choi, Jaesik and Miller, Timothy and Stumpf, Simone and Yang, Guang-Zhong},
  journal   = {Science Robotics},
  title     = {{XAI} -- Explainable artificial intelligence},
  year      = {2019},
  number    = {37},
  pages     = {1--2},
  volume    = {4},
  doi       = {10.1126/scirobotics.aay7120},
  publisher = {American Association for the Advancement of Science ({AAAS})},
}

@Article{arrieta2020explainable,
  author    = {Arrieta, Alejandro Barredo and D{\'{\i}}az-Rodr{\'{\i}}guez, Natalia and Del Ser, Javier and Bennetot, Adrien and Tabik, Siham and Barbado, Alberto and Garcia, Salvador and Gil-Lopez, Sergio and Molina, Daniel and Benjamins, Richard and Chatila, Raja and Herrera, Francisco},
  journal   = {Information Fusion},
  title     = {Explainable Artificial Intelligence ({XAI}): Concepts, taxonomies, opportunities and challenges toward responsible {AI}},
  year      = {2020},
  pages     = {82--115},
  volume    = {58},
  doi       = {10.1016/j.inffus.2019.12.012},
  publisher = {Elsevier},
}

@Article{adadi2018peeking,
  author    = {Adadi, Amina and Berrada, Mohammed},
  journal   = {{IEEE} Access},
  title     = {Peeking Inside the Black-Box: A Survey on Explainable Artificial Intelligence ({XAI})},
  year      = {2018},
  pages     = {52138--52160},
  volume    = {6},
  doi       = {10.1109/access.2018.2870052},
  publisher = {Institute of Electrical and Electronics Engineers ({IEEE})},
}

@Article{langer2021what,
  author    = {Langer, Markus and Oster, Daniel and Speith, Timo and Hermanns, Holger and K\"{a}stner, Lena and Schmidt, Eva and Sesing, Andreas and Baum, Kevin},
  journal   = {Artificial Intelligence},
  title     = {What do we want from Explainable Artificial Intelligence ({XAI})? {\textendash} A stakeholder perspective on {XAI} and a conceptual model guiding interdisciplinary {XAI} research},
  year      = {2021},
  pages     = {1--24},
  volume    = {296},
  doi       = {10.1016/j.artint.2021.103473},
  publisher = {Elsevier},
}

@InProceedings{speith2022review,
  author    = {Speith, Timo},
  booktitle = {Proceedings of the 2022 {ACM} Conference on Fairness, Accountability, and Transparency},
  title     = {A review of taxonomies of explainable artificial intelligence ({XAI}) methods},
  year      = {2022},
  pages     = {2239--2250},
}

@Article{miller2019explanation,
  author    = {Miller, Tim},
  journal   = {Artificial Intelligence},
  title     = {Explanation in Artificial Intelligence: Insights from the social sciences},
  year      = {2019},
  pages     = {1--38},
  volume    = {267},
  doi       = {10.1016/j.artint.2018.07.007},
  publisher = {Elsevier},
}

@Article{wang2019defining,
  author    = {Wang, Pei},
  journal   = {Journal of Artificial General Intelligence},
  title     = {On Defining Artificial Intelligence},
  year      = {2019},
  number    = {2},
  pages     = {1--37},
  volume    = {10},
  doi       = {10.2478/jagi-2019-0002},
  publisher = {De Gruyter},
}

@Article{newell1976computer,
  author  = {Newell, Allen and Simon, Herbert A.},
  journal = {Communications of the {ACM}},
  title   = {Computer Science as Empirical Inquiry: Symbols and Search},
  year    = {1976},
  number  = {3},
  pages   = {113--126},
  volume  = {19},
}

@Article{mccarthy1988mathematical,
  author    = {McCarthy, John},
  journal   = {Daedalus},
  title     = {Mathematical Logic in Artificial Intelligence},
  year      = {1988},
  pages     = {297--311},
  publisher = {JSTOR},
}

@Book{minsky1985society,
  author    = {Minsky, Marvin},
  publisher = {Simon \& Schuster},
  title     = {The Society of Mind},
  year      = {1985},
  address   = {New York, {NY}},
}

@Article{ross2024ai,
  author    = {Ross, Amber},
  journal   = {{AI} \& Society},
  title     = {{AI} and the Expert; A blueprint for the ethical use of opaque {AI}},
  year      = {2024},
  number    = {3},
  pages     = {925--936},
  volume    = {39},
  doi       = {10.1007/s00146-022-01564-2},
  publisher = {Springer},
}

@Article{peters2022explainable,
  author    = {Peters, Uwe},
  journal   = {AI and Ethics},
  title     = {Explainable {AI} Lacks Regulative Reasons: Why {AI} and human decision-making are not equally opaque},
  year      = {2022},
  number    = {3},
  pages     = {1--12},
  volume    = {3},
  doi       = {10.1007/s43681-022-00217-w},
  publisher = {Springer},
}

@PhdThesis{wang1995non,
  author = {Wang, Pei},
  school = {Indiana University},
  title  = {Non-Axiomatic Reason System: Exploring the Essence of Intelligence},
  year   = {1995},
}

@Article{cambria2023survey,
  author    = {Cambria, Erik and Malandri, Lorenzo and Mercorio, Fabio and Mezzanzanica, Mario and Nobani, Navid},
  journal   = {Information Processing \& Management},
  title     = {A Survey on {XAI} and Natural Language Explanations},
  year      = {2023},
  number    = {1},
  pages     = {1--16},
  volume    = {60},
  doi       = {10.1016/j.ipm.2022.103111},
  publisher = {Elsevier},
}

@Article{mamak2023military,
  author    = {Mamak,Kamil and Kowalczewska, Kaja},
  journal   = {Ethics and Information Technology},
  title     = {Military Robots Should Not Look Like Humans},
  year      = {2023},
  number    = {3},
  pages     = {1--10},
  volume    = {25},
  doi       = {10.1007/s10676-023-09718-6},
  publisher = {Springer},
}

@Article{rudin2019stop,
  author    = {Rudin, Cynthia},
  journal   = {Nature Machine Intelligence},
  title     = {Stop explaining black box machine learning models for high stakes decisions and use interpretable models instead},
  year      = {2019},
  number    = {5},
  pages     = {206--215},
  volume    = {1},
  doi       = {10.1038/s42256-019-0048-x},
  publisher = {Nature},
}

@Article{nyholm2017attributing,
  author    = {Nyholm, Sven},
  journal   = {Science and Engineering Ethics},
  title     = {Attributing Agency to Automated Systems: Reflections on Human–Robot Collaborations and Responsibility-Loci},
  year      = {2017},
  number    = {4},
  pages     = {1201--1219},
  volume    = {24},
  doi       = {10.1007/s11948-017-9943-x},
  publisher = {Springer},
}

@Article{trevithick2024everything,
  author  = {Trevithick, Joseph},
  journal = {The Warzone},
  title   = {Everything New We Just Learned About The Collaborative Combat Aircraft Program},
  year    = {2024},
  note    = {\url{https://www.twz.com/air/collaborative-combat-aircraft-poised-to-reshape-the-air-force}},
  url     = {https://www.twz.com/air/collaborative-combat-aircraft-poised-to-reshape-the-air-force},
}

@Article{wood2024explainable,
  author    = {Wood, Nathan Gabriel},
  journal   = {Ethics and Information Technology},
  title     = {Explainable {AI} in the Military Domain},
  year      = {2024},
  number    = {2},
  pages     = {1--13},
  volume    = {26},
  doi       = {10.1007/s10676-024-09762-w},
  publisher = {Springer},
}

@Article{taylor2024explainable,
  author    = {Taylor, Isaac},
  journal   = {AI \& Society},
  title     = {Is Explainable {AI} Responsible {AI}?},
  year      = {2024},
  number    = {3},
  pages     = {1695--1704},
  volume    = {40},
  doi       = {10.1007/s00146-024-01939-7},
  publisher = {Springer},
}

@Article{tsamados2024human,
  author    = {Tsamados, Andreas and Floridi, Luciano and Taddeo, Mariarosaria},
  journal   = {AI and Ethics},
  title     = {Human Control of {AI} Systems: From supervision to teaming},
  year      = {2024},
  doi       = {10.1007/s43681-024-00489-4},
  publisher = {Springer Science and Business Media LLC},
}

@Article{johnson2024finding,
  author    = {Johnson, James},
  journal   = {Global Society},
  title     = {Finding {AI} Faces in the Moon and Armies in the Clouds: Anthropomorphising Artificial Intelligence in Military Human-Machine Interactions},
  year      = {2024},
  number    = {1},
  pages     = {67--82},
  volume    = {38},
  doi       = {10.1080/13600826.2023.2205444},
  publisher = {Informa UK Limited},
}

@Article{wood2024reliability,
  author    = {Wood, Nathan Gabriel},
  journal   = {Ethics and Armed Forces},
  title     = {Reliability Standards for (Autonomous) Weapons},
  year      = {2024},
  number    = {1},
  volume    = {24},
  publisher = {Center for Ethical Education in the Armed Forces - Bundeswehr},
  url       = {https://www.ethikundmilitaer.de/en/magazine-datenbank/detail/01-2024/article/reliability-standards-for-autonomous-weapons-the-enduring-relevance-of-humans},
}

@InCollection{malle2021multidimensional,
  author    = {Malle, Bertram F. and Ullman, Daniel},
  booktitle = {Trust in Human-Robot Interaction},
  publisher = {Academic Press},
  title     = {A Multidimensional Conception and Measure of Human-Robot Trust},
  year      = {2021},
  address   = {London, {UK}},
  editor    = {Nam, Chang S. and Lyons, Joseph B.},
  isbn      = {9780128194720},
  pages     = {3--25},
  doi       = {10.1016/b978-0-12-819472-0.00001-0},
}

@Article{malle2023measuring,
  author    = {Malle, Bertram F. and Ullman, Daniel},
  title     = {Measuring Human-Robot Trust with the {MDMT} (Multi-Dimensional Measure of Trust)},
  year      = {2023},
  pages     = {1--3},
  copyright = {arXiv.org perpetual, non-exclusive license},
  doi       = {10.48550/ARXIV.2311.14887},
  publisher = {arXiv},
}

@Article{barman2024beyond,
  author    = {Barman, Kristian Gonz\'{a}lez and Wood, Nathan Gabriel and Pawlowski, Pawel},
  journal   = {Ethics and Information Technology},
  title     = {Beyond Transparency and Explainability: On the need for adequate and contextualized user guidelines for {LLM} use},
  year      = {2024},
  number    = {3},
  pages     = {1--12},
  volume    = {26},
  doi       = {10.1007/s10676-024-09778-2},
  publisher = {Springer},
}

@TechReport{euhlegai2019ethics,
  author      = {{High-Level Expert Group on Artificial Intelligence}},
  institution = {European Commission},
  title       = {Ethics Guidelines for Trustworthy {AI}},
  year        = {2019},
  publisher   = {European Commission},
}

@TechReport{euparliament2023ai,
  author      = {{European Parliament}},
  institution = {European Parliament},
  title       = {{A}rtificial {I}ntelligence {A}ct},
  year        = {2023},
  publisher   = {European Parliament},
}

@Article{siau2018building,
  author  = {Siau, Keng and Wang, Weiyu},
  journal = {Cutter Business Technology Journal},
  title   = {Building Trust In Artificial Intelligence, Machine Learning, and Robotics},
  year    = {2018},
  number  = {2},
  pages   = {47--53},
  volume  = {31},
}

@InProceedings{ullman2018what,
  author     = {Ullman, Daniel and Malle, Bertram F.},
  booktitle  = {Companion of the 2018 ACM/IEEE International Conference on Human-Robot Interaction},
  title      = {What Does it Mean to Trust a Robot?: Steps Toward a Multidimensional Measure of Trust},
  year       = {2018},
  pages      = {263--264},
  publisher  = {ACM},
  series     = {HRI '18},
  collection = {HRI '18},
  doi        = {10.1145/3173386.3176991},
}

@InProceedings{ullman2019measuring,
  author    = {Ullman, Daniel and Malle, Bertram F.},
  booktitle = {2019 14th ACM/IEEE International Conference on Human-Robot Interaction (HRI)},
  title     = {Measuring Gains and Losses in Human-Robot Trust: Evidence for Differentiable Components of Trust},
  year      = {2019},
  pages     = {618--619},
  publisher = {IEEE},
  doi       = {10.1109/hri.2019.8673154},
}

@InCollection{kwik2023performance,
  author    = {Kwik, Jonathan and van Engers, Tom},
  booktitle = {Artificial Intelligence and Normative Challenges: International and Comparative Legal Perspectives},
  publisher = {Springer},
  title     = {Performance or Explainability? A Law of Armed Conflict Perspective},
  year      = {2023},
  pages     = {255--279},
}

@Book{galliott2015military,
  author    = {Galliott, Jai},
  publisher = {Ashgate},
  title     = {Military Robots},
  year      = {2015},
  address   = {Farnham, {UK}},
  isbn      = {9781472426628},
  series    = {Military and Defence Ethics},
  subtitle  = {Mapping the moral landscape},
}

@Book{johnson2023ai,
  author    = {Johnson, James},
  publisher = {Oxford University Press},
  title     = {The {AI} Commander: Centaur Teaming, Command, and Ethical Dilemmas},
  year      = {2023},
  address   = {Oxford, {UK}},
  isbn      = {9780198892267},
}

@Article{scharre2016centaur,
  author    = {Scharre, Paul},
  journal   = {Temple International and Comparative Law Journal},
  title     = {Centaur Warfighting: The false choice of humans vs. automation},
  year      = {2016},
  pages     = {151--165},
  volume    = {30},
  publisher = {Hein Online},
}

@TechReport{macak2005centaurs,
  author      = {Macak, Christopher Andrew},
  institution = {United States Marine Corps Command and Staff College},
  title       = {Centaurs for Maneuver Warfare: Human-Machine Collaboration and Manned-Unmanned Teaming for the Fifth-Generation Ground Combat Element},
  year        = {2005},
}

@TechReport{neads2021tools,
  author      = {Neads, Alex and Galbreath, David J. and Farrell, Theo},
  institution = {Australian Army Research Centre},
  title       = {From Tools to Teammates: Human-Machine Teaming and the Future of Command and Control in the {A}ustralian {A}rmy},
  year        = {2021},
}

@Article{hadi2023large,
  author    = {Hadi, Muhammad Usman and Qureshi, Rizwan and Shah, Abbas and Irfan, Muhammad and Zafar, Anas and Shaikh, Muhammad Bilal and Akhtar, Naveed and Wu, Jia and Mirjalili, Seyedali},
  journal   = {Authorea Preprints},
  title     = {Large Language Models: A comprehensive survey of its applications, challenges, limitations, and future prospects},
  year      = {2023},
  doi       = {10.36227/techrxiv.23589741.v4},
  publisher = {Institute of Electrical and Electronics Engineers (IEEE)},
}

@Article{boiko2023emergent,
  author    = {Boiko, Daniil A. and MacKnight, Robert and Gomes, Gabe},
  journal   = {arXiv Preprint},
  title     = {Emergent Autonomous Scientific Research Capabilities of Large Language Models},
  year      = {2023},
  doi       = {10.48550/ARXIV.2304.05332},
  publisher = {arXiv},
}

@Article{fan2023bibliometric,
  author    = {Fan, Lizhou and Li, Lingyao and Ma, Zihui and Lee, Sanggyu and Yu, Huizi and Hemphill, Libby},
  journal   = {arXiv Preprint},
  title     = {A Bibliometric Review of Large Language Models Research from 2017 to 2023},
  year      = {2023},
  doi       = {10.48550/ARXIV.2304.02020},
  publisher = {arXiv},
}

@Article{inagaki2023llms,
  author    = {Inagaki, Takashi and Kato, Akari and Takahashi, Koichi and Ozaki, Haruka and Kanda, Genki N.},
  journal   = {arXiv Preprint},
  title     = {{LLMs} Can Generate Robotic Scripts From Goal-Oriented Instructions in Biological Laboratory Automation},
  year      = {2023},
  doi       = {10.48550/ARXIV.2304.10267},
  publisher = {arXiv},
}

@Article{williams2023algorithmic,
  author    = {Williams, Nigel and Ivanov, Stanislav and Buhalis, Dimitrios},
  journal   = {arXiv Preprint},
  title     = {Algorithmic Ghost in the Research Shell: Large Language Models and Academic Knowledge Creation in Management Research},
  year      = {2023},
  doi       = {10.48550/ARXIV.2303.07304},
  publisher = {arXiv},
}

@Article{rajpurkar2022ai,
  author    = {Rajpurkar, Pranav and Chen, Emma and Banerjee, Oishi and Topol, Eric J},
  journal   = {Nature Medicine},
  title     = {{AI} in Health and Medicine},
  year      = {2022},
  number    = {1},
  pages     = {31--38},
  volume    = {28},
  doi       = {https://doi.org/10.1038/s41591-021-01614-0},
  publisher = {Nature},
}

@Article{johnson2021precision,
  author    = {Johnson, Kevin B. and Wei, Wei-Qi and Weeraratne, Dilhan and Frisse, Mark E. and Misulis, Karl and Rhee, Kyu and Zhao, Juan and Snowdon, Jane L.},
  journal   = {Clinical and Translational Science},
  title     = {Precision Medicine, {AI}, and the Future of Personalized Health Care},
  year      = {2021},
  number    = {1},
  pages     = {86--93},
  volume    = {14},
  doi       = {10.1111/cts.12884},
  publisher = {Wiley},
}

@Article{king2023future,
  author    = {King, Michael R.},
  journal   = {Annals of Biomedical Engineering},
  title     = {The Future of {AI} in Medicine: A perspective from a Chatbot},
  year      = {2023},
  number    = {2},
  pages     = {291--295},
  volume    = {51},
  doi       = {10.1007/s10439-022-03121-w},
  publisher = {Springer},
}

@Article{stahl2018ethics,
  author    = {Stahl, Bernd Carsten and Wright, David},
  journal   = {IEEE Security \& Privacy},
  title     = {Ethics and Privacy in {AI} and Big Data: Implementing responsible research and innovation},
  year      = {2018},
  number    = {3},
  pages     = {26--33},
  volume    = {16},
  doi       = {10.1109/msp.2018.2701164},
  publisher = {Institute of Electrical and Electronics Engineers (IEEE)},
}

@Article{elliott2022ai,
  author    = {Elliott, David and Soifer, Eldon},
  journal   = {Frontiers in Artificial Intelligence},
  title     = {AI Technologies, Privacy, and Security},
  year      = {2022},
  volume    = {5},
  doi       = {10.3389/frai.2022.826737},
  publisher = {Frontiers},
}

@Article{xiang2024fairness,
  author    = {Xiang, Alice},
  journal   = {Science and Technology Law Review},
  title     = {Fairness \& Privacy in an Age of Generative {AI}},
  year      = {2024},
  number    = {2},
  volume    = {25},
  doi       = {10.52214/stlr.v25i2.12765},
  publisher = {Columbia University Libraries},
}

@Article{morley2020ethics,
  author    = {Morley, Jessica and Machado, Caio C.V. and Burr, Christopher and Cowls, Josh and Joshi, Indra and Taddeo, Mariarosaria and Floridi, Luciano},
  journal   = {Social Science \& Medicine},
  title     = {The Ethics of {AI} in Health Care: A mapping review},
  year      = {2020},
  volume    = {260},
  doi       = {10.1016/j.socscimed.2020.113172},
  publisher = {Elsevier},
}

@Article{gundersen2022future,
  author    = {Gundersen, Torbj\o{}rn and B\ae{}r\o{}e, Kristine},
  journal   = {Science and Engineering Ethics},
  title     = {The Future Ethics of Artificial Intelligence in Medicine: Making Sense of Collaborative Models},
  year      = {2022},
  number    = {2},
  volume    = {28},
  doi       = {10.1007/s11948-022-00369-2},
  publisher = {Springer},
}

@Article{anom2020ethics,
  author    = {Anom, B.Y.},
  journal   = {Ethics, Medicine and Public Health},
  title     = {Ethics of Big Data and Artificial Intelligence in Medicine},
  year      = {2020},
  pages     = {1--11},
  volume    = {15},
  doi       = {10.1016/j.jemep.2020.100568},
  publisher = {Elsevier},
}

@Article{soni2020artificial,
  author    = {Soni, Neha and Sharma, Enakshi Khular and Singh, Narotam and Kapoor, Amita},
  journal   = {Procedia Computer Science},
  title     = {Artificial Intelligence in Business: From Research and Innovation to Market Deployment},
  year      = {2020},
  volume    = {167},
  doi       = {10.1016/j.procs.2020.03.272},
  publisher = {Elsevier},
}

@InProceedings{mattmann2024ai,
  author       = {Mattmann, Chris A. and Broderick, Daniel},
  booktitle    = {2024 IEEE Aerospace Conference},
  title        = {{AI} Hype Versus Reality -- Will It Work for You?},
  year         = {2024},
  organization = {IEEE},
  pages        = {1--10},
  doi          = {10.1109/AERO58975.2024},
}

@Article{ryan2020trust,
  author    = {Ryan, Mark},
  journal   = {Science and Engineering Ethics},
  title     = {In {AI} We Trust: Ethics, Artificial Intelligence, and Reliability},
  year      = {2020},
  number    = {5},
  pages     = {2749--2767},
  volume    = {26},
  doi       = {10.1007/s11948-020-00228-y},
  publisher = {Springer},
}

@Article{omrani2022trust,
  author    = {Omrani, Nessrine and Rivieccio, Giorgia and Fiore, Ugo and Schiavone, Francesco and Agreda, Sergio Garcia},
  journal   = {Technological Forecasting and Social Change},
  title     = {To trust or not to trust? An assessment of trust in {AI}-based systems: Concerns, ethics and contexts},
  year      = {2022},
  pages     = {121763},
  volume    = {181},
  doi       = {10.1016/j.techfore.2022.121763},
  publisher = {Elsevier},
}

@Article{choung2022trust,
  author    = {Choung, Hyesun and David, Prabu and Ross, Arun},
  journal   = {International Journal of Human–Computer Interaction},
  title     = {Trust in {AI} and Its Role in the Acceptance of {AI} Technologies},
  year      = {2022},
  number    = {9},
  pages     = {1727--1739},
  volume    = {39},
  doi       = {10.1080/10447318.2022.2050543},
  publisher = {Informa UK Limited},
}

@InCollection{lyons2019trust,
  author    = {Lyons, Joseph B. and Wynne, Kevin T. and Mahoney, Sean and Roebke, Mark A.},
  booktitle = {Artificial Intelligence for the Internet of Everything},
  publisher = {Elsevier},
  title     = {Trust and Human-Machine Teaming: A Qualitative Study},
  year      = {2019},
  editor    = {Lawless, William and Mitu, Ranjeev and Sofge, Donald and Moskowitz, Ira S. and Russell, Stuart},
  pages     = {101--116},
  doi       = {10.1016/b978-0-12-817636-8.00006-5},
}

@Article{oneill2020human,
  author    = {O'Neill, Thomas and McNeese, Nathan and Barron, Amy and Schelble, Beau},
  journal   = {Human Factors: The Journal of the Human Factors and Ergonomics Society},
  title     = {Human-Autonomy Teaming: A Review and Analysis of the Empirical Literature},
  year      = {2020},
  number    = {5},
  pages     = {904--938},
  volume    = {64},
  doi       = {10.1177/0018720820960865},
  publisher = {SAGE},
}

@InCollection{baier2014trust,
  author    = {Baier, Annette},
  booktitle = {Feminist Social Thought},
  publisher = {Routledge},
  title     = {Trust and Antitrust},
  year      = {2014},
  address   = {New York, {NY}},
  editor    = {Meyers, Diana Tietjens},
  isbn      = {9781135025021},
  pages     = {604--629},
  doi       = {10.4324/9780203705841},
}

@Book{hardin2002trust,
  author    = {Hardin, Russell},
  publisher = {Russell Sage Foundation},
  title     = {Trust and Trustworthiness},
  year      = {2002},
  address   = {New York, {NY}},
}

@Article{mayer1995integrative,
  author    = {Mayer, Roger C. and Davis, James H. and Schoorman, F. David},
  journal   = {The Academy of Management Review},
  title     = {An Integrative Model of Organizational Trust},
  year      = {1995},
  number    = {3},
  pages     = {709},
  volume    = {20},
  doi       = {10.2307/258792},
  publisher = {Academy of Management},
}

@Article{koller1988risk,
  author    = {Koller, Michael},
  journal   = {Basic and Applied Social Psychology},
  title     = {Risk as a Determinant of Trust},
  year      = {1988},
  number    = {4},
  pages     = {265--276},
  volume    = {9},
  doi       = {10.1207/s15324834basp0904_2},
  publisher = {Informa UK Limited},
}

@InProceedings{josang2004analysing,
  author    = {J\o{}sang, Audun and Presti, St\'{e}phane Lo},
  booktitle = {Trust Management: Second International Conference, iTrust 2004},
  title     = {Analysing the Relationship between Risk and Trust},
  year      = {2004},
  address   = {Berlin, Germany},
  editor    = {Jensen, Christian and Poslad, Stefan and Dimitrakos, Theo},
  pages     = {135--145},
  publisher = {Springer},
  doi       = {10.1007/978-3-540-24747-0_11},
  isbn      = {9783540247470},
}

@Article{mckinlay2020trust,
  author    = {McKinlay, Steve},
  journal   = {Big Data and Democracy},
  title     = {Trust and Algorithmic Opacity},
  year      = {2020},
  pages     = {153--166},
  address   = {Edinburgh, {UK}},
  publisher = {Edinburgh University Press},
}

@Article{kumar2024opacity,
  author    = {Kumar, Manohar and Aijaz, Aisha and Chattar, Omkar and Shukla, Jainendra and Mutharaju, Raghava},
  journal   = {IEEE Transactions on Affective Computing},
  title     = {Opacity, Transparency, and the Ethics of Affective Computing},
  year      = {2024},
  number    = {1},
  pages     = {4--17},
  volume    = {15},
  doi       = {10.1109/taffc.2023.3278230},
  publisher = {Institute of Electrical and Electronics Engineers (IEEE)},
}

@Article{saghiri2022survey,
  author    = {Saghiri, Ali Mohammad and Vahidipour, S. Mehdi and Jabbarpour, Mohammad Reza and Sookhak, Mehdi and Forestiero, Agostino},
  journal   = {Applied Sciences},
  title     = {A Survey of Artificial Intelligence Challenges: Analyzing the Definitions, Relationships, and Evolutions},
  year      = {2022},
  number    = {8},
  pages     = {4054},
  volume    = {12},
  doi       = {10.3390/app12084054},
  publisher = {MDPI},
}

@Article{duan2022survey,
  author    = {Duan, Jiafei and Yu, Samson and Tan, Hui Li and Zhu, Hongyuan and Tan, Cheston},
  journal   = {IEEE Transactions on Emerging Topics in Computational Intelligence},
  title     = {A Survey of Embodied {AI}: From Simulators to Research Tasks},
  year      = {2022},
  number    = {2},
  pages     = {230--244},
  volume    = {6},
  doi       = {10.1109/tetci.2022.3141105},
  publisher = {Institute of Electrical and Electronics Engineers (IEEE)},
}

@Article{li2023trustworthy,
  author    = {Li, Bo and Qi, Peng and Liu, Bo and Di, Shuai and Liu, Jingen and Pei, Jiquan and Yi, Jinfeng and Zhou, Bowen},
  journal   = {ACM Computing Surveys},
  title     = {Trustworthy {AI}: From Principles to Practices},
  year      = {2023},
  number    = {9},
  pages     = {1--46},
  volume    = {55},
  doi       = {10.1145/3555803},
  publisher = {Association for Computing Machinery (ACM)},
}

@InProceedings{lakkaraju2020how,
  author     = {Lakkaraju, Himabindu and Bastani, Osbert},
  booktitle  = {Proceedings of the AAAI/ACM Conference on AI, Ethics, and Society},
  title      = {``{H}ow do I fool you?'': Manipulating User Trust via Misleading Black Box Explanations},
  year       = {2020},
  publisher  = {ACM},
  series     = {AIES '20},
  collection = {AIES '20},
  doi        = {10.1145/3375627.3375833},
}

@Article{zhang2019why,
  author    = {Zhang, Yujia and Song, Kuangyan and Sun, Yiming and Tan, Sarah and Udell, Madeleine},
  title     = {``{W}hy Should You Trust My Explanation?'' Understanding Uncertainty in {LIME} Explanations},
  year      = {2019},
  doi       = {10.48550/ARXIV.1904.12991},
  publisher = {arXiv},
}

@Book{waller1958dogs,
  author    = {Waller, Anna M.},
  publisher = {Department of the Army, Office of the Quartermaster General},
  title     = {Dogs and National Defense},
  year      = {1958},
}

@PhdThesis{hammerstrom2005ground,
  author = {Hammerstrom, Michael L.},
  school = {Naval Postgraduate School},
  title  = {Ground Dog Day Lessons Don't Have to be Relearned in the Use of Dogs in Combat},
  year   = {2005},
}

@Article{lee2024why,
  author    = {Lee, Jin Hwa and Mannelli, Stefano Sarao and Saxe, Andrew},
  title     = {Why Do Animals Need Shaping? A Theory of Task Composition and Curriculum Learning},
  year      = {2024},
  doi       = {10.48550/ARXIV.2402.18361},
  publisher = {arXiv},
}

@Article{briefer2024goats,
  author    = {Briefer, Elodie F. and Haque, Samaah and Baciadonna, Luigi and McElligott, Alan G.},
  journal   = {Frontiers in Zoology},
  title     = {Goats Excel at Learning and Remembering a Highly Novel Cognitive Task},
  year      = {2014},
  number    = {1},
  pages     = {20},
  volume    = {11},
  doi       = {10.1186/1742-9994-11-20},
  publisher = {Springer},
}

@Article{anderson2001teams,
  author    = {Anderson, C.},
  journal   = {Behavioral Ecology},
  title     = {Teams in animal societies},
  year      = {2001},
  number    = {5},
  pages     = {534--540},
  volume    = {12},
  doi       = {10.1093/beheco/12.5.534},
  publisher = {Oxford University Press},
}

@Article{cruse2014military,
  author    = {Cruse, Sarah D.},
  journal   = {Animal Law},
  title     = {Military Working Dogs: Classification and treatment in the {US} Armed Forces},
  year      = {2014},
  pages     = {249--284},
  volume    = {21},
  publisher = {Hein Online},
}

@Article{demirbas2016adults,
  author    = {Demirbas, Yasemin Salgirli and Ozturk, Hakan and Emre, Bahri and Kockaya, Mustafa and Ozvardar, Tarkan and Scott, Alison},
  journal   = {Anthrozo\"{o}s},
  title     = {Adults' Ability to Interpret Canine Body Language during a Dog-Child Interaction},
  year      = {2016},
  number    = {4},
  pages     = {581--596},
  volume    = {29},
  doi       = {10.1080/08927936.2016.1228750},
  publisher = {Informa UK Limited},
}

@Article{demirtas2023dog,
  author    = {Demirtas, Ahu and Atilgan, Durmus and Saral, Begum and Isparta, Sevim and Ozturk, Hakan and Ozvardar, Tarkan and Demirbas, Yasemin Salgirli},
  journal   = {Journal of Veterinary Behavior},
  title     = {Dog Owners' Recognition of Pain-Related Behavioral Changes in Their Dogs},
  year      = {2023},
  pages     = {39--46},
  volume    = {62},
  doi       = {10.1016/j.jveb.2023.02.006},
  publisher = {Elsevier BV},
}

@Article{salles2020anthropomorphism,
  author    = {Salles, Arleen and Evers, Kathinka and Farisco, Michele},
  journal   = {AJOB Neuroscience},
  title     = {Anthropomorphism in {AI}},
  year      = {2020},
  number    = {2},
  pages     = {88--95},
  volume    = {11},
  doi       = {10.1080/21507740.2020.1740350},
  publisher = {Informa UK Limited},
}

@PhdThesis{simmons2010humanizing,
  author    = {Simmons, Eugene Bradley},
  title     = {Humanizing the Humvee: Personification techniques and visual rhetoric as used in a {US} Army technical comic book},
  year      = {2010},
  publisher = {Iowa State University},
}

@MastersThesis{anderson2020bomber,
  author = {Anderson, Sarah Marie},
  school = {University of Georgia},
  title  = {Bomber Boys and Their ``Girls'': Intimate Bonds Formed Between Eighth Air Force Airmen and Bomber Aircraft During the Second World War as Reflected in Nose Art},
  year   = {2020},
}

@Article{cappuccio2021saving,
  author    = {Cappuccio, Massimiliano L. and Galliott, Jai C. and Sandoval, Eduardo B.},
  journal   = {International Journal of Social Robotics},
  title     = {Saving Private Robot: Risks and Advantages of Anthropomorphism in Agent-Soldier Teams},
  year      = {2021},
  number    = {10},
  pages     = {2135--2148},
  volume    = {14},
  doi       = {10.1007/s12369-021-00755-z},
  publisher = {Springer},
}

@Article{dorsch2024explainable,
  author    = {Dorsch, John and Moll, Maximilian},
  journal   = {{arXiv}},
  title     = {Explainable and Human-Grounded {AI} for Decision Support Systems: The Theory of Epistemic Quasi-Partnerships},
  year      = {2024},
  doi       = {10.48550/ARXIV.2409.14839},
  publisher = {arXiv},
}

@Article{sparrow2007killer,
  author    = {Sparrow, Robert},
  journal   = {Journal of Applied Philosophy},
  title     = {Killer Robots},
  year      = {2007},
  number    = {1},
  pages     = {62--77},
  volume    = {24},
  doi       = {10.1111/j.1468-5930.2007.00346.x},
  publisher = {Wiley},
}

@Article{sparrow2016robots,
  author    = {Sparrow, Robert},
  journal   = {Ethics \& International Affairs},
  title     = {Robots and Respect: Assessing the Case Against Autonomous Weapon Systems},
  year      = {2016},
  number    = {01},
  pages     = {93--116},
  volume    = {30},
  doi       = {https://doi.org/10.1017/s0892679415000647},
  publisher = {Cambridge University Press},
}

@Article{evans2023collaborate,
  author    = {Evans, Katie D. and Robbins, Scott A. and Bryson, Joanna J.},
  journal   = {Topics in Cognitive Science},
  title     = {Do We Collaborate With What We Design?},
  year      = {2023},
  doi       = {10.1111/tops.12682},
  publisher = {Wiley},
}

@InCollection{sullins2011robot,
  author    = {Sullins, John P.},
  booktitle = {Machine Ethics},
  publisher = {Cambridge University Press},
  title     = {When Is a Robot a Moral Agent},
  year      = {2011},
  address   = {Cambridge, {UK}},
  editor    = {Anderson, Michael and Anderson, Susan Leigh},
  isbn      = {978-0-521-11235-2},
  number    = {2001},
  pages     = {151--161},
  volume    = {6},
}

@Article{antoniadi2021current,
  author    = {Antoniadi, Anna Markella and Du, Yuhan and Guendouz, Yasmine and Wei, Lan and Mazo, Claudia and Becker, Brett A. and Mooney, Catherine},
  journal   = {Applied Sciences},
  title     = {Current Challenges and Future Opportunities for {XAI} in Machine Learning-Based Clinical Decision Support Systems: A Systematic Review},
  year      = {2021},
  number    = {11},
  pages     = {1--23},
  volume    = {11},
  doi       = {https://doi.org/10.3390/app11115088},
  publisher = {MDPI},
}

@Article{machlev2022explainable,
  author    = {Machlev, R. and Heistrene, L. and Perl, M. and Levy, K.Y. and Belikov, J. and Mannor, S. and Levron, Y.},
  journal   = {Energy and {AI}},
  title     = {Explainable Artificial Intelligence ({XAI}) techniques for energy and power systems: Review, challenges and opportunities},
  year      = {2022},
  pages     = {1--13},
  volume    = {9},
  doi       = {https://doi.org/10.1016/j.egyai.2022.100169},
  publisher = {Elsevier},
}

@Article{kalasampath2025review,
  author    = {Kalasampath, Khushi and Spoorthi, K. N. and Sajeev, Sreeparvathy and Kuppa, Sahil Sarma and Ajay, Kavya and Maruthamuthu, Angulakshmi},
  journal   = {IEEE Access},
  title     = {A Literature Review on Applications of Explainable Artificial Intelligence ({XAI})},
  year      = {2025},
  pages     = {41111--41140},
  volume    = {13},
  doi       = {https://doi.org/10.1109/ACCESS.2025.3546681},
  publisher = {Institute of Electrical and Electronics Engineers (IEEE)},
}

@Article{hsu2009real,
  author     = {Hsu, Jeremy},
  journal    = {{NBC} News},
  title      = {Real Soldiers Love Their Robot Brethren},
  year       = {2009},
  note       = {\url{https://www.nbcnews.com/id/wbna30868033}},
  readstatus = {read},
  url        = {https://www.nbcnews.com/id/wbna30868033},
}

@Article{seffers2023man,
  author     = {Seffers, George I.},
  journal    = {Signal},
  title      = {Ge. {J}ames {R}ainey: Man-machine integration may revolutionize combat arms},
  year       = {2023},
  note       = {\url{https://www.afcea.org/signal-media/gen-james-rainey-man-machine-integration-may-revolutionize-combat-arms#:~:text=The%20integration%20of%20human%20beings,commander%2C%20U.S.%20Army%20Futures%20Command.}},
  readstatus = {read},
  url        = {https://www.afcea.org/signal-media/gen-james-rainey-man-machine-integration-may-revolutionize-combat-arms#:~:text=The%20integration%20of%20human%20beings,commander%2C%20U.S.%20Army%20Futures%20Command.},
}

@Article{park2024mastering,
  author     = {Park, Allyson},
  journal    = {National Defense},
  title      = {Mastering Human-Machine Learning Cricial for {A}ir {F}orce, Official Says},
  year       = {2024},
  note       = {\url{https://www.nationaldefensemagazine.org/articles/2024/2/28/just-in-mastering-humanmachine-learning-crucial-for-air-force-official-says}},
  readstatus = {read},
  url        = {https://www.nationaldefensemagazine.org/articles/2024/2/28/just-in-mastering-humanmachine-learning-crucial-for-air-force-official-says},
}

\end{document}